\pdfoutput=1
\documentclass[a4paper,fleqn,usenatbib]{mnras}
\usepackage{graphicx}	
\usepackage{amsmath}	
\usepackage{amssymb}	
\usepackage{bm}
\usepackage{txfonts}

\usepackage[T1]{fontenc}
\usepackage{ae,aecompl}
\usepackage{hyperref}
\usepackage{color}

\title[Evolution of circumbinary accretion disk around SMBBH: post-Newtonian HD versus Newtonian HD]{Evolution of circumbinary accretion disk around supermassive binary black hole: post-Newtonian hydrodynamics versus Newtonian hydrodynamics}

\author[Wenshuai Liu]{
Wenshuai Liu\thanks{E-mail: 674602871@qq.com (DPF)}\\
School of Physics, Huazhong University of Science and Technology, Wuhan 430074, China\\
}

\date{Accepted XXX. Received YYY; in original form ZZZ}

\begin{document}
\label{firstpage}
\pagerange{\pageref{firstpage}--\pageref{lastpage}}
\maketitle

\begin{abstract}
We study the evolution of accretion disk around a supermassive binary black hole with equal mass using non-relativistic hydrodynamical simulations performed with FARGO3D. Compared with previous studies with the Newtonian hydrodynamics, here, we adopt the post-Newtonian hydrodynamics using the near zone metric of the binary black hole. In contrast to the Newtonian investigation, we find that there is a dramatic difference in the post-Newtonian regime, gap formed by the circumbinary accretion disk around the binary with equal mass is wider with the post-Newtonian hydrodynamics than that with the Newtonian hydrodynamics and is independent of disk viscosity given that hydrodynamical simulations are run for about the same factor times the viscous timescale associated with different viscosities. This may present unique observable signatures of the continuum emission in such binary-disk system.
\end{abstract}




\section{Introduction}

There is clear evidence that centers of all massive galaxies in the local Universe harbour a supermassive black hole with mass between $10^6$M$_{\odot}$ and $10^9$M$_{\odot}$\citep{Kormendy95,Ferrarese05,Kormendy13}. Furthermore, supermassive binary black hole (SMBBH) are expected to form in the merged galaxy remnant after a galaxy-galaxy merger according to the current hierarchical formation models\citep{White78}. Then the interaction of the newly formed SMBBH with the disk of gas and stars in the merged galaxy brings the SMBBH close enough to become a bound SMBBH. However, SMBBH have to evolve from distance of hundreds of kpc to sub-pc scale so as to merge quickly well within a Hubble time due to the emission of gravitational radiation which will be detected by the proposed space-based laser interferometers and the ongoing International Pulsar Timing Array \citep{Hobbs10}.

After the merger of two galaxies, dynamical friction from stars and gas will take the SMBBH to the center of the merged galaxy, forming a wide gravitationally bound binary with a semi-major axis of tens of pc \citep{Milosavljevic01}. However, interaction of the binary with individual stars depletes its loss cone when the binary begins to harden, resulting in ineffective dynamical friction and stalling the SMBBH at pc scale known as the 'final pc problem' \citep{Begelman80,Milosavljevic03,Merritt05}. To overcome this bottleneck, a viable mechanism is proposed that the interaction of the SMBBH with the surrounding accretion disk is able to further extract angular momentum and energy of the binary until the inspiral stage dominated by gravitational radiation \citep{Armitage05,MacFadyen08,Cuadra09,Lodato09,Haiman09} since most SMBBHs are expected to coalesce in a gas-rich environment at the center of the newly formed galaxy \citep{Cuadra09,Chapon13,Colpi14} which provides an ideal target for gravitational wave observations and electromagnetic counterparts.

Large numbers of works have extensively studied the binary-disk system in the Newtonian regime with analytic methods \citep{Paczynski77,Papaloizou77,Kocsis12a,Kocsis12b} and hydrodynamical simulations \citep{Lin79,Artymowicz94,MacFadyen08,Mayama10,de ValBorro11,Shi12,D'Orazio13,Farris14,Munoz16,Nelson16,Ju16,Miranda17,Tang17,Tang18,Munoz19}. Simulations within the Newtonian regime during the inspiral of the binary black hole have also been investigated \citep{Baruteau12,Cerioli16,Farris15,Tang18}. Two-dimensional simulations by \citet{Noble12} and three-dimensional ones by \citet{Zilhao15} conducted the fully relativistic magnetohydrodynamic evolution of a circumbinary disk surrounding two non-spinning black holes with equal mass using the near zone metric. Later on, the fully relativistic magnetohydrodynamic simulations of SMBBH-disk interaction with both the near zone metric and the inner zone metric have been performed \citep{Bowen17,Bowen18,d'Ascoli18,Bowen19} and these simulations are limited to several binary orbits. In particular, \citet{Bowen17} showed that the gravitational potential of the binary black hole in the post-Newtonian regime is shallower than that in the Newtonian regime, and that such shallow potential in the post-Newtonian regime has significant effects on the dynamics of the mini-disk around each SMBH. Simulations of SMBBH-disk interaction with full numerical relativity are so computationally expensive that the evolution of disk around SMBBH is constrained within the stage near merger \citep{Farris12,Giacomazzo12}. In this work, we present hydrodynamical simulations of accretion of equal mass SMBBH by using post-Newtonian (PN) hydrodynamics in which the near zone metric of SMBBH is used.

The PN approximation is effective in describing the spacetime evolution of physical systems in which motions are slow compared with the speed of light and gravitational fields are relatively weak, solving the Einstein field equations perturbatively. PN theory has also recently been used to interface with numerical relativity simulations to serve as the initial data for BBH mergers \citep{Tichy03,Bonning03,Yunes07,Kelly07,Campanelli09,Kelly10,Mundim11}. In the case of SMBBH with sufficiently small separation where the motion of black hole is slow [$(v/c)^2 \ll 1$] and the effect of general relativity emerges, the PN approximation could provide an adequate description of the spacetime far from the SMBBH better than Newtonian theory.

Due to the fact that PN approximation is the expansion of Einstein field equations perturbatively, simulations with PN hydrodynamics could be faster than that with full numerical relativity, allowing simulations of evolution of disk around SMBBH with a long time. That is to say, simulations of interaction of SMBBH of small separation with the surrounding disk by using PN hydrodynamics have the advantages that PN is more accurate than Newtonian and is faster than the full numerical relativity. Hydrodynamical simulations in the Newtonian regime could describe the binary-disk interaction precisely, provided the separation of the binary is relatively large. When the binary separation becomes small enough combined with the condition that the binary still needs a long time to merge, simulations of the evolution of such system could be precise and fast by PN hydrodynamics. Thus, PN hydrodynamics could be adopted as the method to simulate the transitional stage between the stage where Newtonian regime could account for the binary-disk evolution when the separation of the binary is sufficiently large and the one where the full numerical relativity should be used in order to simulate the binary-disk interaction precisely when the separation of the binary is sufficiently small.

Independent of the gravitational wave emitted by SMBBH during inspiral, interaction of the SMBBH with the surrounding accretion disk could provide peculiar electromagnetic signatures. Simulational results from many studies \citep{MacFadyen08,Miranda17,Munoz16,Munoz19,Shi12,Tang17,Tang18} demonstrated a gap could form during the binary-disk interaction, showing unique observable signatures of the continuum emission \citep{Gultekin12,D'Orazio13,Yan15,Ryan17,Tang18}. Circumbinary disk in the PN regime and in the Newtonian regime may present different property of the continuum emission when the separation of SMBBH is so small that correction from general relativity should be taken into account. PN may also influence the luminosity since the luminosity of the circumbinary disk is proportional to its surface density and such investigation is beyond the scope of this work and will be studied in future.

For the first time, this work investigates the binary-disk interaction with the PN hydrodynamics. As for the simulated region, we concentrate on the
near zone, which is sufficiently far from the horizons so that the weak-field approximation of PN is valid while less than a reduced gravitational wave wavelength away from the center of mass of the binary. The near zone metric is not valid near each SMBH, so we restrict the region of simulation to be larger than a given value. We find that effects of PN hydrodynamics could produce qualitative difference from Newtonian hydrodynamics. Gap around the binary with equal mass is wider when adopting the PN hydrodynamics. Furthermore, simulations with PN hydrodynamics with different viscosities give rise to a wider gap around the binary provided that simulations are run for about the same factor times the viscous timescale associated with different viscosities. Such wider gaps resulting from PN hydrodynamics with different viscosities show similar properties, meaning that PN effect shows up at all viscosities and is independent of disk viscosity. Throughout this work, we use the metric signature (-,+,+,+). We introduce the physics of PN hydrodynamics with the near zone spacetime of SMBBH and give the initial condition used in our simulations in Section 2. We present the results in Section 3. Conclusions are in Section 4.

\section{POST-NEWTONIAN HYDRODYNAMICS WITH INITIAL CONDITION}
We study the tidal interaction between a SMBBH of equal mass and the gaseous disk it is embedded in by using the Newtonian hydrodynamics and the PN hydrodynamics with the near zone metric for comparison to illustrate the importance of the PN effect. To account for the PN effect in our simulations, we use 1PN expansion of the metric of two non-spinning particles for the PN hydrodynamics \citep{Blanchet98,Blanchet14} and fix the orbit of the binary as the way done in \citet{Zilhao15}.

With the fixed separation of the binary, the other question is the PN dynamics accounting for the accretion disk in the region of the near zone of the SMBBH. At the level of approximation of the near zone , we get the dynamics for a test particle
\begin{equation}
\frac{d^2 x^\mu}{d\tau^2}+{\Gamma^\mu}_{\alpha \beta} \frac{dx^\alpha}{d\tau} \frac{dx^\beta}{d\tau}=0 \label{1}
\end{equation}
where
\begin{equation}
{\Gamma^\mu}_{\alpha \beta} = \frac{1}{2} g^{\mu \sigma}\left( \partial_\alpha g_{\beta \sigma} + \partial_\beta g_{\alpha \sigma}  - \partial_\sigma g_{\nu \kappa} \right)\label{2}
\end{equation}
\noindent where $g_{\alpha \beta}$ is given in \citet{Blanchet14} or \citet{Alvi00} in detail. In this work, we adopt 1PN expansion of $g_{\alpha \beta}$ from the near zone metric as follows

\begin{eqnarray}
g_{tt} & = & -1+ {2m_1 \over r_1} + {2m_2 \over r_2}-2({m_1 \over r_1}+{m_2 \over r_2})^2
\nonumber \\
     & + &		{m_1 \over r_1}\left[4v_1^2 - ({\bf n}_1\cdot{\bf v}_1)^2\right] + {m_2 \over r_2}\left[4v_2^2 - ({\bf n}_2\cdot{\bf v}_2)^2\right]
 \nonumber \\
    & + & m_1 m_2(-\frac{r_1}{2b^3}-\frac{r_2}{2b^3}+ \frac{r_1^2}{2r_2b^3}
 \nonumber \\
    & + & \frac{r_2^2}{2r_1b^3}-\frac{5}{2r_2b}-\frac{5}{2r_1b}),\\ \label{3}
	g_{0i} &=& -4\left({m_1 \over r_1} v_1^i +
			{m_2 \over r_2} v_2^i \right),\\ \label{4}
	g_{ij} &=& \delta_i{}_j \left(1 + {2m_1 \over r_1} +
			{2m_2 \over r_2} \right).\label{5}	
\end{eqnarray}

\noindent where $G$ and $c$ are set to 1. $\alpha, \beta = 0,1,2,3$. $m_1$ and $m_2$ represent the two black holes mass, $r_1$ and $r_2$ the distance between a field point and the two black holes, $v_1$ and $v_2$ the velocity of the two black holes relative to the center of mass. b is the separation of the two black holes. ${\bf n}_1$ and ${\bf n}_2$ are the normalized position vectors.

The PN configuration will match the Newtonian one in the limit of a large binary separation according to the metric above. When the separation b is large enough, the circular velocity $v_1$ and $v_2$ of the two black holes are sufficiently small. Combined with the condition that the distance between a fluid element and the two black holes, $r_1$ and $r_2$ is large enough, the metric above returns to the Newtonian regime. Thus, in the case of a large binary separation, results of simulations with PN hydrodynamics will be consistent with that of Newtonian hydrodynamics, which will be confirmed by the following simulations.

To investigate the gravitational interaction between SMBBH and the gaseous disk, effect of the disk on the evolution of the binary black hole is neglected. With the geodesic equation, dynamics of the fluid element in the gaseous disk under the 1PN metric of the SMBBH is given as follows
\begin{equation}
\frac{d^2 x^i}{d\tau^2}=-{\Gamma^i}_{\alpha \beta} \frac{dx^\alpha}{d\tau} \frac{dx^\beta}{d\tau}
\end{equation}
where $i=1,2,3$\\
compared with
\begin{equation}
\frac{d^2 x^i}{dt^2}=\frac{\partial \Phi}{\partial x^i}
\end{equation}
in the Newtonian regime where $\Phi$ is the Newtonian potential.

With equations (\ref{1})-(\ref{4}),we obtain the following equation up to $O(c^{-2})$
\begin{eqnarray}\label{PN_geo}
\frac{dv^j}{dt}&=&\partial_jU+\frac{1}{c^2}\Big[(v^2-4U)\partial_jU\\
&-&(4v^k\partial_kU+3\partial_tU)v^j\\
&-&4v^k\left(\partial_jU_k-\partial_kU_j\right)+4\partial_tU_j+\partial_j\Psi]
\end{eqnarray}
where
\begin{equation}
U={m_1 \over r_1} + {m_2 \over r_2}
\end{equation}
\begin{equation}
U_i={m_1 \over r_1} v_1^i +{m_2 \over r_2} v_2^i  \label{6}
\end{equation}

\begin{eqnarray}
\Psi &=&{m_1 \over 2r_1}\left[4v_1^2 - ({\bf n}_1\cdot{\bf v}_1)^2\right] + {m_2 \over 2r_2}\left[4v_2^2 - ({\bf n}_2\cdot{\bf v}_2)^2\right]
 \nonumber \\
    & + & \frac{m_1 m_2}{2}(-\frac{r_1}{2b^3}-\frac{r_2}{2b^3}+ \frac{r_1^2}{2r_2b^3}
 \nonumber \\
    & + & \frac{r_2^2}{2r_1b^3}-\frac{5}{2r_2b}-\frac{5}{2r_1b})
\end{eqnarray}

Similar to the method in \citet{Poisson14}, \citet{Nazari17} and \citet{Nazari18}, we get the 1PN hydrodynamical equations as follows

\begin{eqnarray}\label{conti_eq}
\frac{d\rho^*}{dt}&=&-\rho^*\nabla\cdot \bm{v}\\
\rho^*\frac{d\bm{v}}{dt}&=&\rho^*\nabla U-\nabla p+\frac{1}{c^2}\Big\lbrace\big(\frac{v^2}{2}+U
\nonumber \\
&+&\Pi+\frac{p}{\rho^*}\big)\nabla p-\bm{v}\frac{\partial p}{\partial t}+\rho^*\Big[\left(v^2-4U\right)\nabla U
\nonumber \\
&-&4\bm{v}\times\left(\nabla\times\boldsymbol{\bm{U}}\right)+4\frac{\partial \boldsymbol{\bm{U}}}{\partial t}-\bm{v}\big(3\frac{\partial U}{\partial t}+4\bm{v}\cdot\nabla U\big)
\nonumber \\
&+&\nabla\Psi]\Big\rbrace+\nabla \cdot \bm{T}\\
\rho^*\frac{d\Pi}{dt}&=&\frac{p}{\rho^*}\frac{d\rho^*}{dt}
\end{eqnarray}
where $\boldsymbol{\bm{U}}$ is defined in equation (\ref{6}). $d/ dt=\partial/\partial t+\bm{v}\cdot\nabla$, $\rho^*=\rho\left(1+v^2/2c^2+3U/c^2\right)$, $\Pi$ is the internal energy per unit mass. Since viscosity in general relativity is an open problem \citep{Bemfica18}, we adopt the tensor $\bm{T} = \rho^* \nu \left[ \nabla \bm{v} + \left(\nabla \bm{v} \right)^T - \frac{2}{3}\left(\nabla \cdot \bm{v}\right) \bm{I} \right]$ for simplicity. $\bm{I}$ is the unit tensor of the same rank as $\nabla \bm{v}$.

The Newtonian hydrodynamical equations used for simulations are

\begin{eqnarray}\label{conti_eq}
\frac{d\rho}{dt}&=&-\rho\nabla\cdot \bm{v}\\
\rho\frac{d\bm{v}}{dt}&=&\rho\nabla U-\nabla p+\nabla \cdot \bm{T}\\
\rho\frac{d\Pi}{dt}&=&\frac{p}{\rho}\frac{d\rho}{dt}
\end{eqnarray}

With the above equations and the equation of state $p=c_s^2\rho$, we have a set of equations which account for the fluid dynamics in 1PN and Newtonian regime.

Before performing the simulations, one needs to generate the initial disk near quasiequilibrium both in Newtonian and PN regime. Starting with a disk which is in quasiequilibrium by following the similar procedure described in \citet{Kazemi18} and generating the disk with circular orbit under the azimuthal averaged metric, we have the circumbinary disk in quasiequilibrium around the SMBBH. The resulting velocity in the azimuthal direction is

\begin{eqnarray}
v_{\varphi} & \simeq &  \sqrt{F}+\frac{1}{c^2}\bigg\lbrace 2(R\,U_{\varphi })'+\frac{R}{\sqrt{F}}\Big(\frac{1}{2} R \left(U'\right)^2+\, (U^2)'
\nonumber \\
&-& \frac{p'}{4\, \rho^* } (\frac{R\, p'}{\rho^* }+\frac{2\, p}{\rho^* }+2\, \Pi +R\,U'+2\,U)-\frac{\Psi '}{2}\Big)\bigg\rbrace \label{PNv1}
\end{eqnarray}
where $F=\frac{R\,p'}{ \rho^* }-R\,U'$, $\rho^*=\rho\left(1+v^2/2c^2+3U/c^2\right)$, $'=\frac{\partial}{\partial R}$ and $U_{\varphi }$, $U$, $U^2$, $\Psi$ are averaged in the azimuthal direction.

The corresponding angular velocity along with radius is

\begin{align}
\Omega\simeq\frac{v_{\varphi}}{R} \label{PNo1}
\end{align}

The above two equations are in the PN regime. In the Newtonian regime, these two equations could be written as

\begin{align}
v_{\varphi}&\simeq  \sqrt{F} \label{PNv2}
\end{align}
and
\begin{align}
\Omega\simeq\frac{\sqrt{F}}{R} \label{PNo2}
\end{align}
where $F=\frac{R\,p'}{ \rho }-R\,U'$

The algorithm of the modified FARGO3D is similar to the original FARGO3D except that the source term is changed from $\rho\nabla U-\nabla p+\nabla \cdot \bm{T}$ to

\begin{eqnarray}
&&\rho^*\nabla U-\nabla p+\frac{1}{c^2}\Big\lbrace\big(\frac{v^2}{2}+U
\nonumber \\
&+&\Pi+\frac{p}{\rho^*}\big)\nabla p-\bm{v}\frac{\partial p}{\partial t}+\rho^*\Big[\left(v^2-4U\right)\nabla U
\nonumber \\
&-&4\bm{v}\times\left(\nabla\times\bm{U}\right)+4\frac{\partial \bm{U}}{\partial t}-\bm{v}\big(3\frac{\partial U}{\partial t}+4\bm{v}\cdot\nabla U\big)
\nonumber \\
&+&\nabla\Psi]\Big\rbrace+\nabla \cdot \bm{T}
\end{eqnarray}

The Newtonian and PN hydrodynamical simulations have the same initial gaseous density distribution. At the beginning of the simulations with PN hydrodynamics, we generate $\rho^*$ from the equation $\rho^*=\rho\left(1+v^2/2c^2+3U/c^2\right)$. During each timestep, we update $\rho^*$ in the PN hydrodynamical simulations and generate the output of $\rho$ from the equation $\rho^*=\rho\left(1+v^2/2c^2+3U/c^2\right)$ and the output is generated after every binary orbit.

We carry out 2D hydrodynamical simulations with FARGO3D \citep{Ben16,Masset00}, and adopt a cylindrical coordinate system centered onto the center of mass of the SMBBH. The mass of each black hole is $M_1 =5\times10^8 M_{\odot}$ in our simulations. We simulate the binary-disk interaction with two sets of separation of the binary $a_0 =40 r_{\rm s}$ and $a_0 =80 r_{\rm s}$ where $r_{\rm s} = 2GM_1 / c^2$. Code unit with $G=1$, $M_1=1$, and $\frac{a_0}{2}=1$ is used and the disk torque acting on the binary is neglected. The near zone metric we use is accurate at distance larger than $r_i \approx (\frac{a^4 M^2_i}{M})^\frac{1}{5}$ from the black hole with mass $M_i$ \citep{Yunes06,Yunes106,JohnsonMcDaniel09} where $i=1,2$ and $M=M_1+M_2$, we find that the near zone metric is effective at distance larger than $r\approx15r_{\rm s}$ from each black hole for $a_0 =40 r_{\rm s}$ and $r\approx26r_{\rm s}$ for $a_0 =80 r_{\rm s}$, so the inner boundary $r_{\rm in}$ should be $r_{\rm in}>\frac{a_0}{2}+15r_s=35r_s=1.75$ for $a_0 =40 r_{\rm s}$ and $r_{\rm in}>\frac{a_0}{2}+26r_s=66r_s=1.65$ for $a_0 =80 r_{\rm s}$ in code unit. Thus, we set the inner boundary at $r_{\rm in}=1.8$ in code unit, the outer boundary is at $r_{\rm out}=10.8$.

The disk is assumed to be locally isothermal with an aspect ratio $h(r)=H/r=0.08\times (r/r_0)^{-0.25}$ where $r_0=a_0/2$. Previous studies \citep{MacFadyen08,Miranda17,Munoz16,Munoz19,Shi12,Tang17,Tang18} on the interaction of equal mass binary with circumbinary disk have shown that a gap forms within about $2.5a$ where $a$ is the separation of the binary, so we set the inner radius of the initial gaseous material disk to be $3a$ ($r=6$ in code unit). The surface density of the disk is modelled as $\Sigma=\Sigma_0 (\frac{r}{r_0})^{-\delta}$ where $\delta=0.5$ and the initial disk has a total mass of about 0.003$M_1$ extending from 6 to 10.8 relative to the center of mass of the binary. The equation of state of the locally isothermal disk is

\begin{equation}
p=c_s^2(r)\Sigma
\end{equation}
where $c_s(r)=hr\Omega$ is the sound speed and $\Omega$ is given in equation (\ref{PNo1}) and (\ref{PNo2}) in the PN and Newtonian regime, respectively.

The viscous gas has a uniform kinematic viscosity $\nu_1=10^{-5}$ ($\alpha\approx 0.0015$)and $\nu_2=2\times10^{-4}$ ($\alpha\approx 0.03$) in order to study the effect of different value of the uniform kinematic viscosity on the evolution of the accretion disk. We perform simulations on an equally spaced grid with the radial grid $N_r=500$ and $N_\phi=500$ covering the full $2\pi$ in azimuth. In the simulation code, the azimuthal boundary condition is periodic due to the full $2\pi$ span. For the radial boundary, we set the inner boundary which allows material to get out of but not into the simulated region ($v_r$ is reset to be zero if $v_r>0$ at the inner boundary) and the outer boundary which doesn't permit material to get into or out of the simulated region ($v_r$ is reset to be zero at the outer boundary) in order to track the accretion flow onto the inner boundary.

Due to the condition that the near zone metric is valid only when the distance to each black hole is larger than $r_i \approx (\frac{a^4 M^2_i}{M})^\frac{1}{5}$ and that the code FARGO3D we use has a finite computational domain with the inner edge and the outer edge, thus, the inner region surrounding the binary black hole is excised from the computational domain in this work, neglecting a possible result that materials entering into and getting out of the inner region after interacting with the binary may affect the size and shape of the gap of the circumbinary disk when the inner region is taken into consideration. In future, we will derive the PN hydrodynamics with the inner zone metric which accounts for the dynamics of material near each black hole besides the near zone metric, and conduct the simulations with hydrodynamical code which could excise the horizon of each black hole and enable the dynamics of material near each black hole.

\begin{figure*}
   \begin{center}
     \begin{tabular}{cc}
     \includegraphics[width=0.33\textwidth]{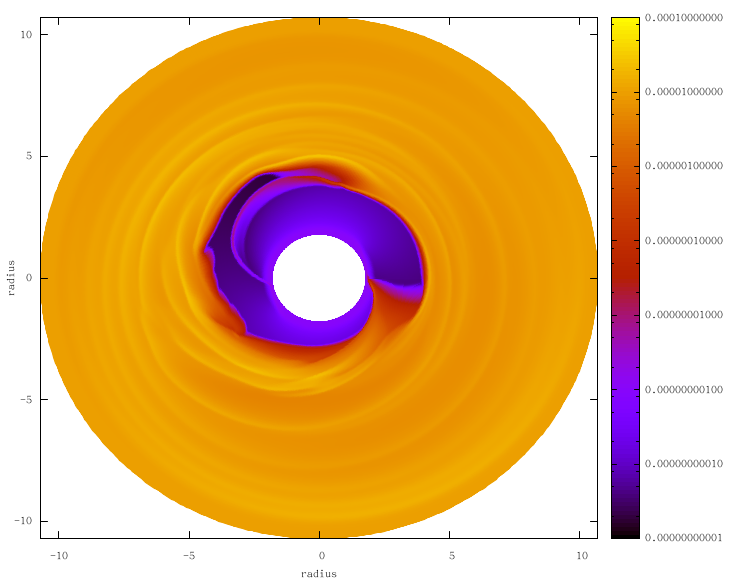}
     \includegraphics[width=0.33\textwidth]{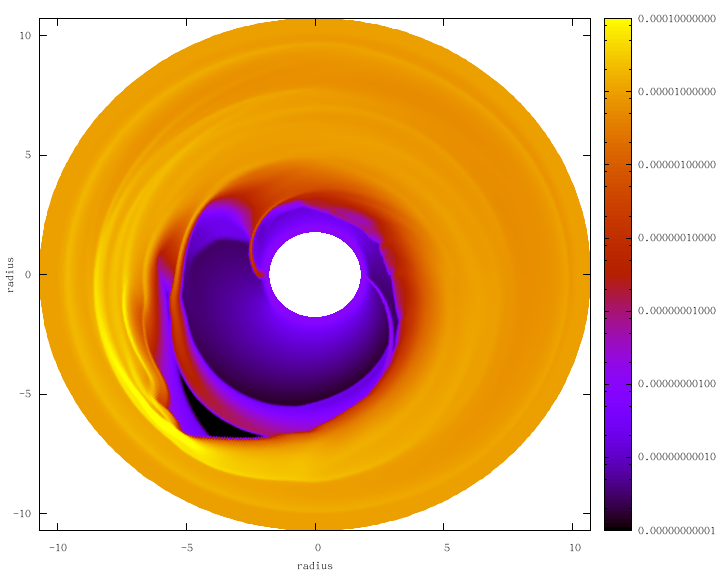}
     \includegraphics[width=0.33\textwidth]{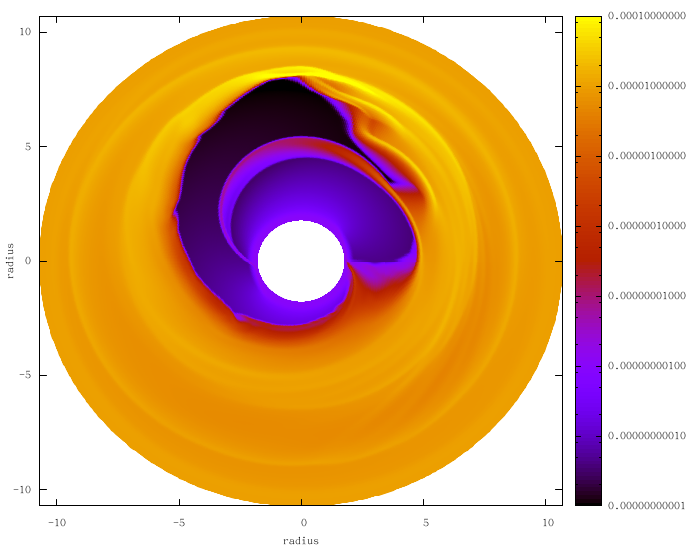}
     \end{tabular}
   \end{center}

    \caption{Left and middle show the surface density after about 500 orbits of the binary with the Newtonian and post-Newtonian hydrodynamics when $a_0 =40 r_{\rm s}$ with $\nu=2\times10^{-4}$, respectively. Right shows the results with the post-Newtonian hydrodynamics when $a_0 =80 r_{\rm s}$ with $\nu=2\times10^{-4}$ (the results with the Newtonian hydrodynamics when $a_0 =80 r_{\rm s}$ with $\nu=2\times10^{-4}$ is identical to the left panel and isn't displayed here).  }
    \label{fig:figure1}
\end{figure*}

\begin{figure*}
 \begin{center}
   \begin{tabular}{cc}
      \includegraphics[width=0.33\textwidth]{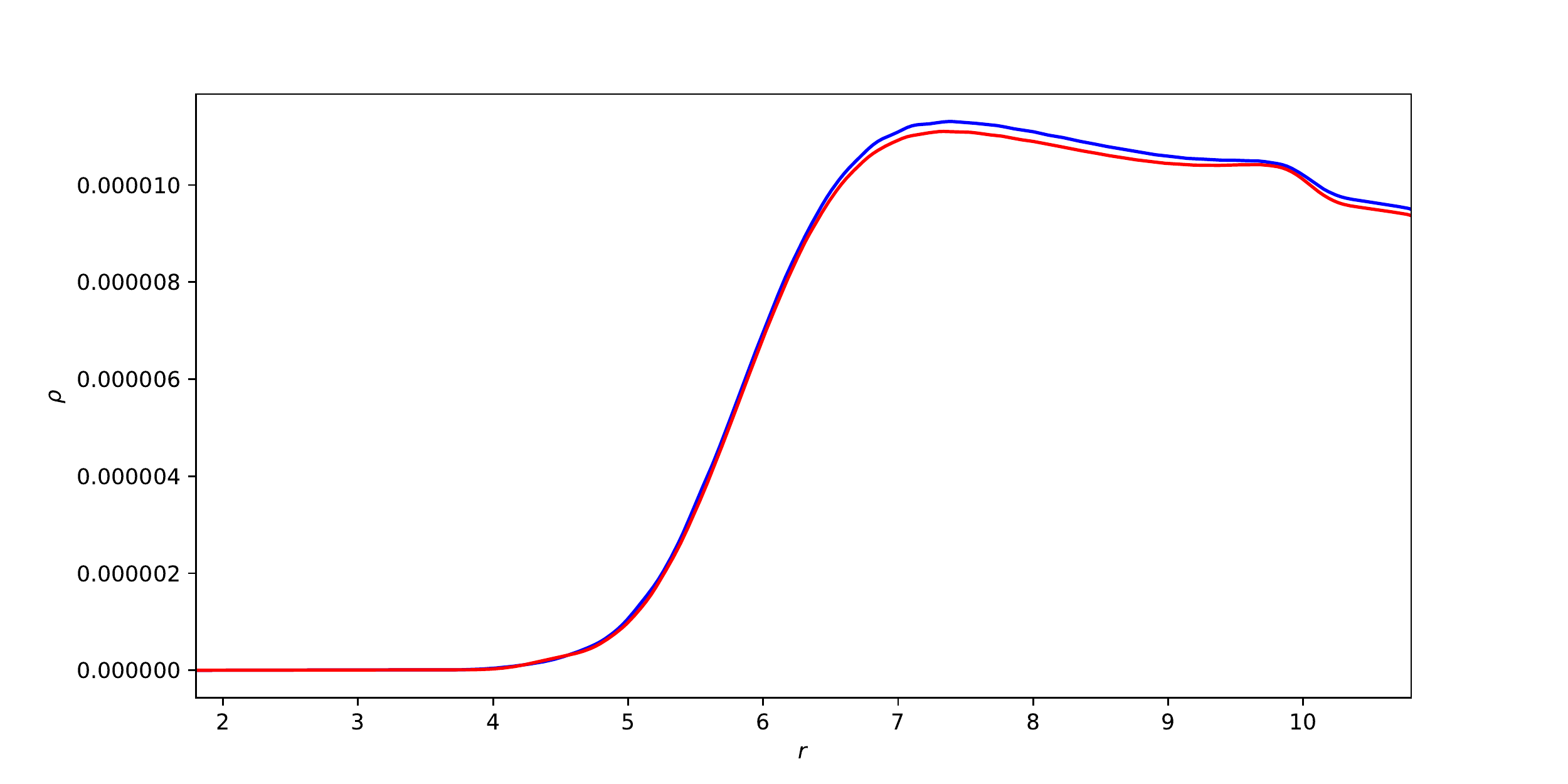}
      \includegraphics[width=0.33\textwidth]{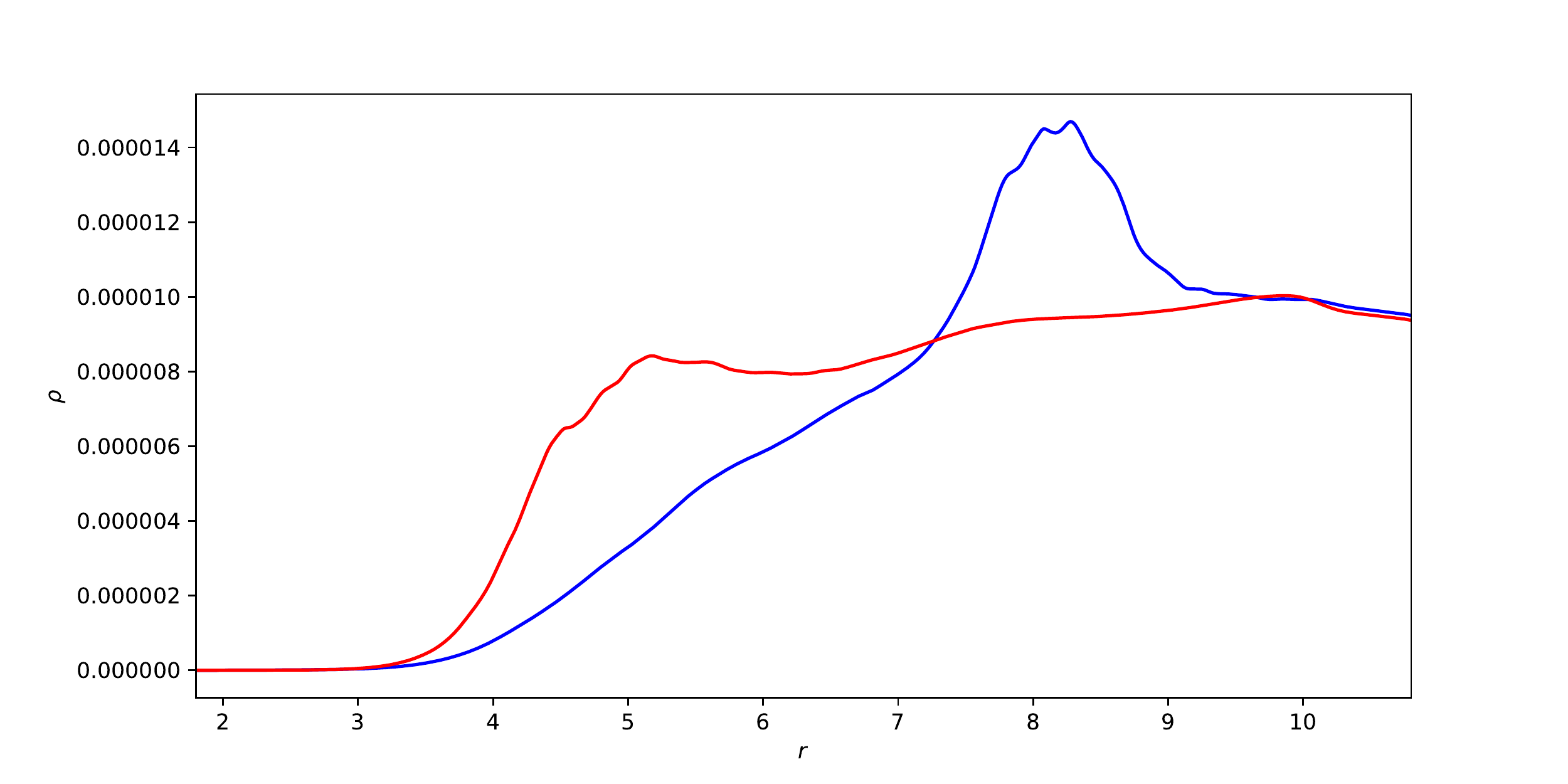}
      \includegraphics[width=0.33\textwidth]{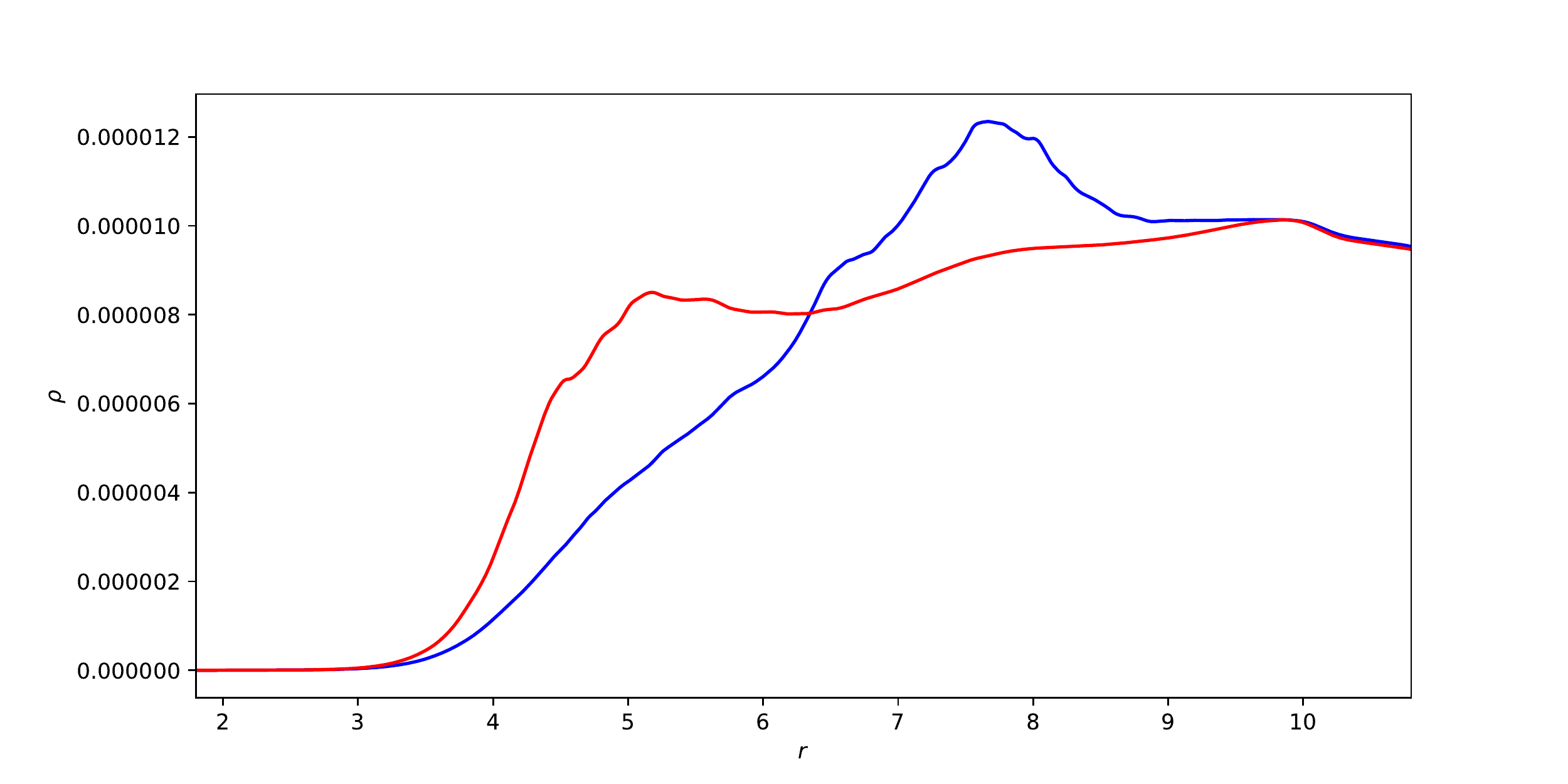}
   \end{tabular}
 \end{center}
    \caption {The azimuthally averaged surface density along with radius which is averaged over 100 orbits from $400$ to $500$ orbits of the binary with the Newtonian (red line) and post-Newtonian (blue line) hydrodynamics. Left and middle show the results when $\nu=10^{-5}$ and $\nu=2\times10^{-4}$ with $a_0 =40 r_{\rm s}$, respectively. Right shows the results when $\nu=2\times10^{-4}$ with $a_0 =80 r_{\rm s}$. }
    \label{fig:figure2}
\end{figure*}

\begin{figure*}
 \begin{center}
   \begin{tabular}{cc}
      \includegraphics[width=0.33\textwidth]{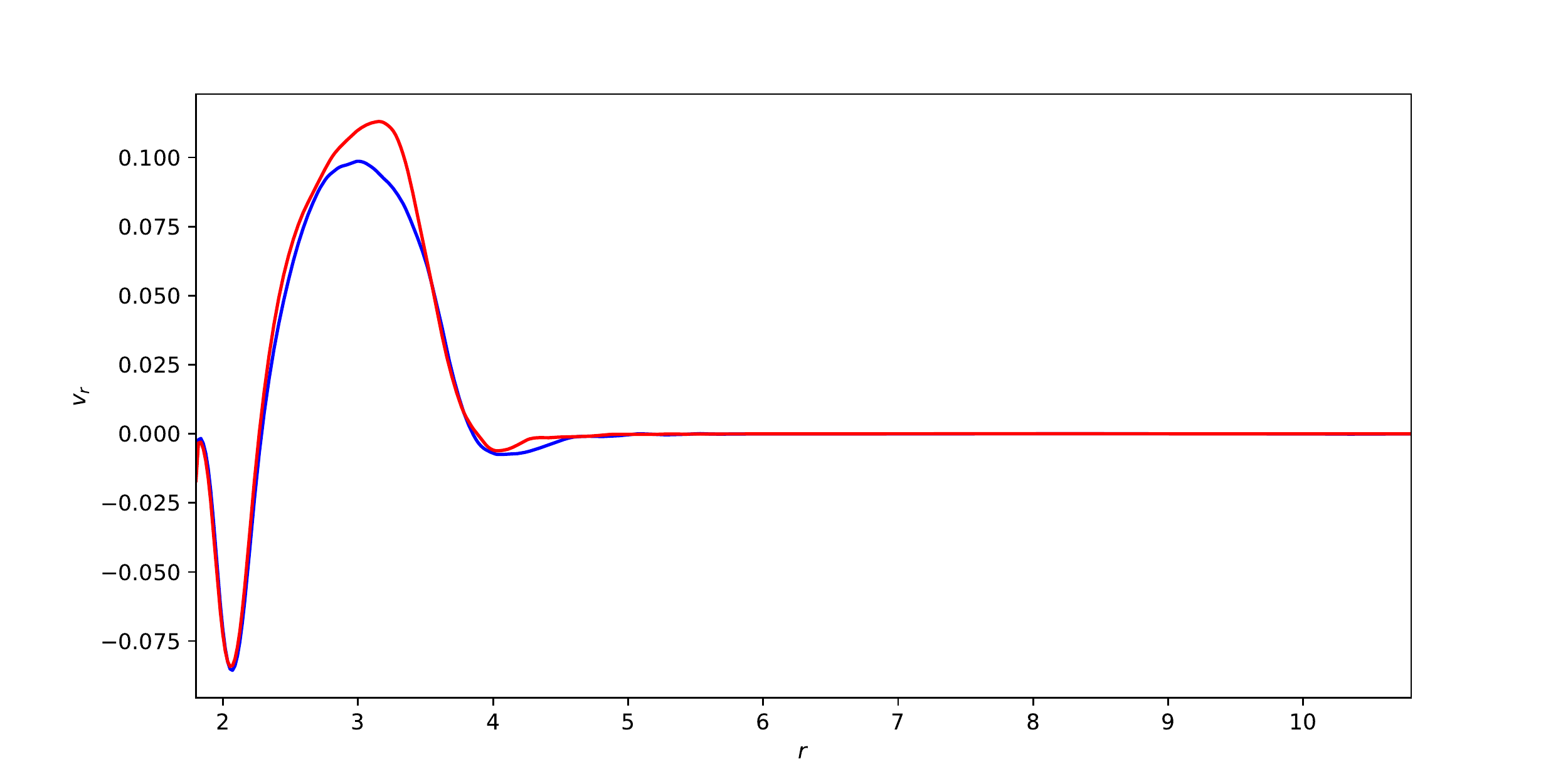}
      \includegraphics[width=0.33\textwidth]{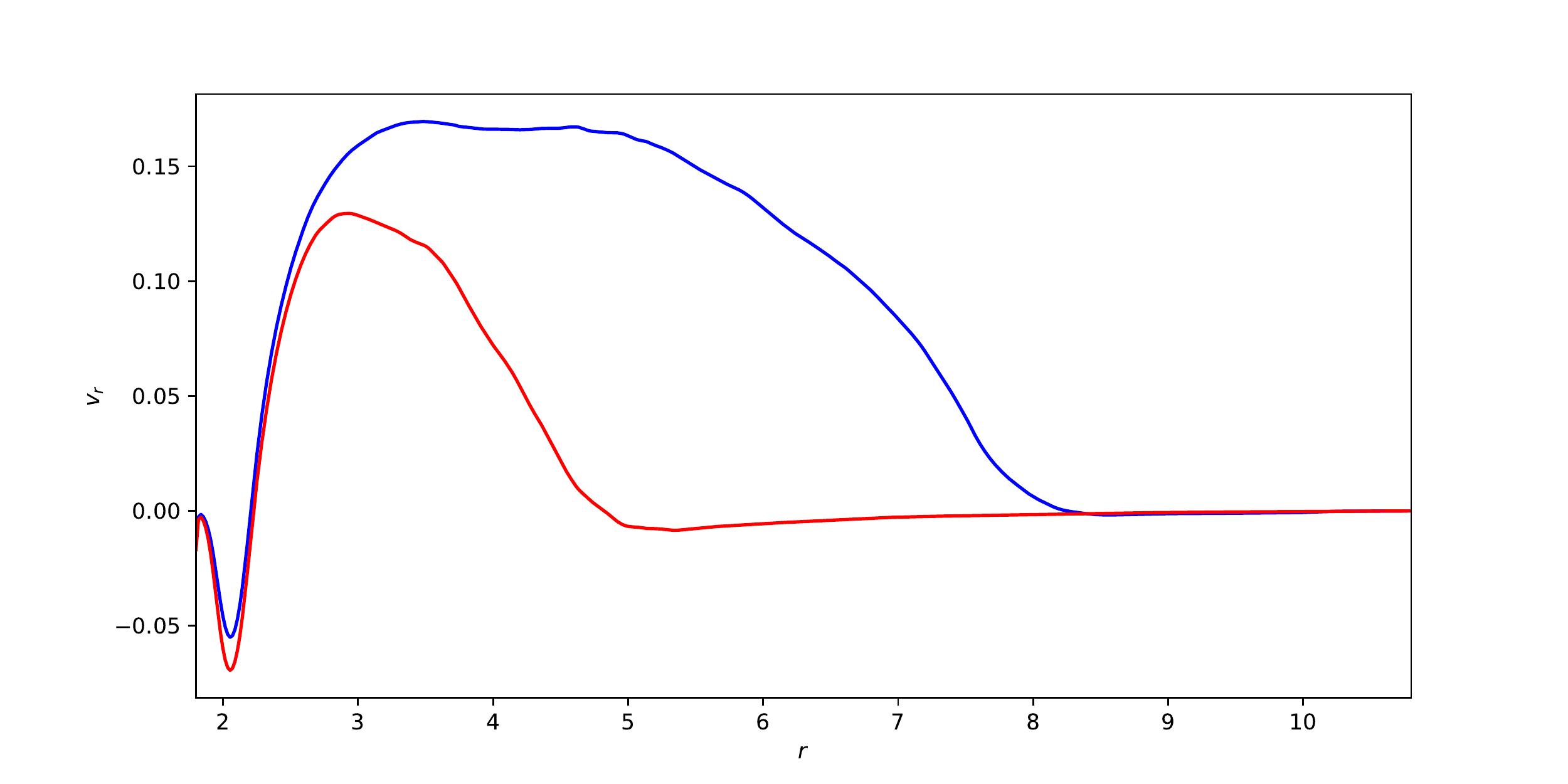}
      \includegraphics[width=0.33\textwidth]{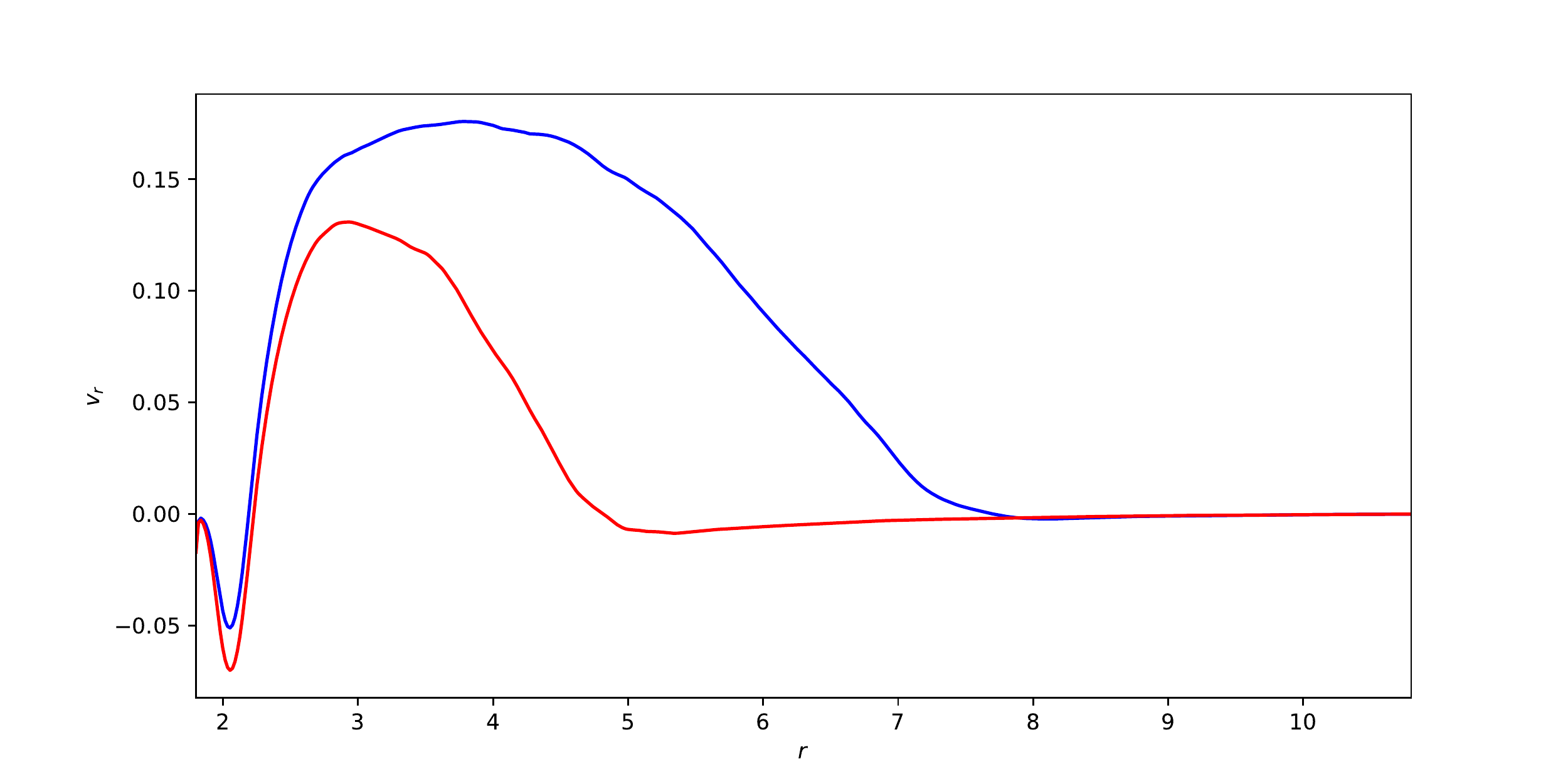}
   \end{tabular}
 \end{center}
    \caption{The azimuthally averaged radial velocity along with radius which is averaged over 100 orbits from $400$ to $500$ orbits of the binary with the Newtonian (red line) and post-Newtonian (blue line) hydrodynamics. Left and middle show the results when $\nu=10^{-5}$ and $\nu=2\times10^{-4}$ with $a_0 =40 r_{\rm s}$, respectively. Right shows the results when $\nu=2\times10^{-4}$ with $a_0 =80 r_{\rm s}$.}
    \label{fig:figure3}
\end{figure*}

\begin{figure*}
 \begin{center}
   \begin{tabular}{cc}
      \includegraphics[width=0.33\textwidth]{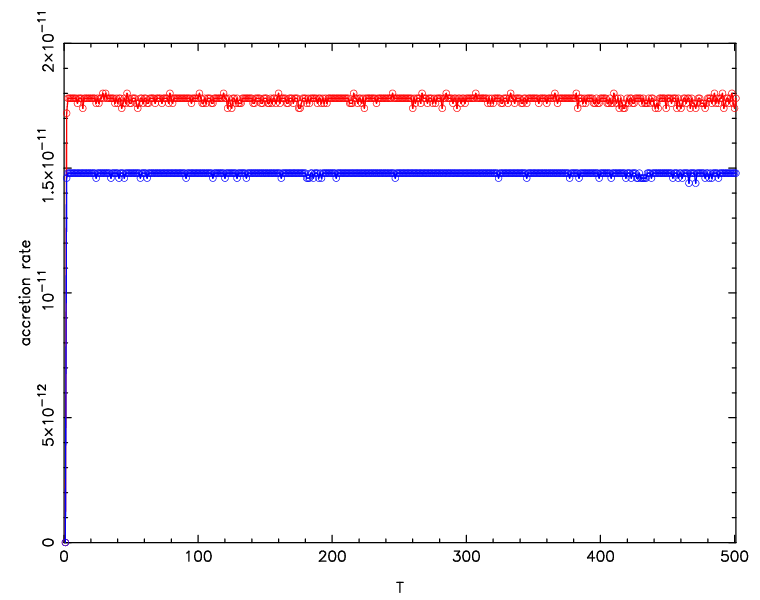}
      \includegraphics[width=0.33\textwidth]{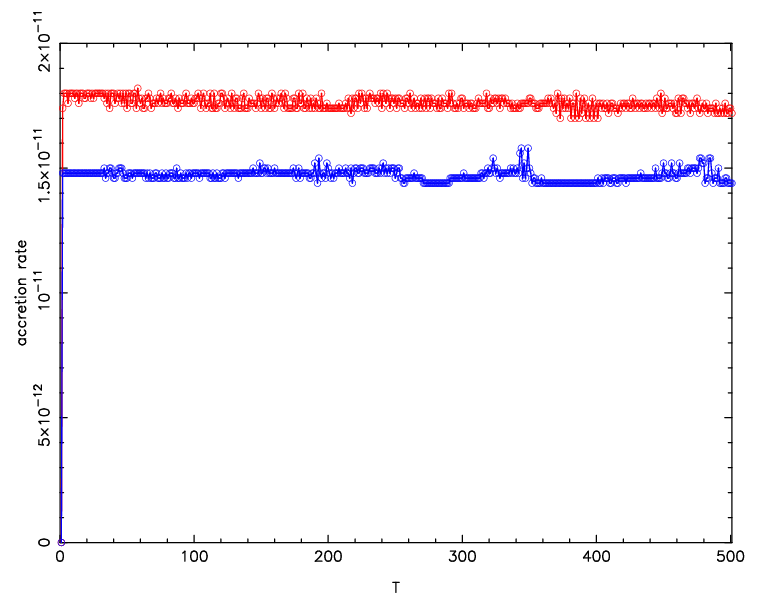}
      \includegraphics[width=0.33\textwidth]{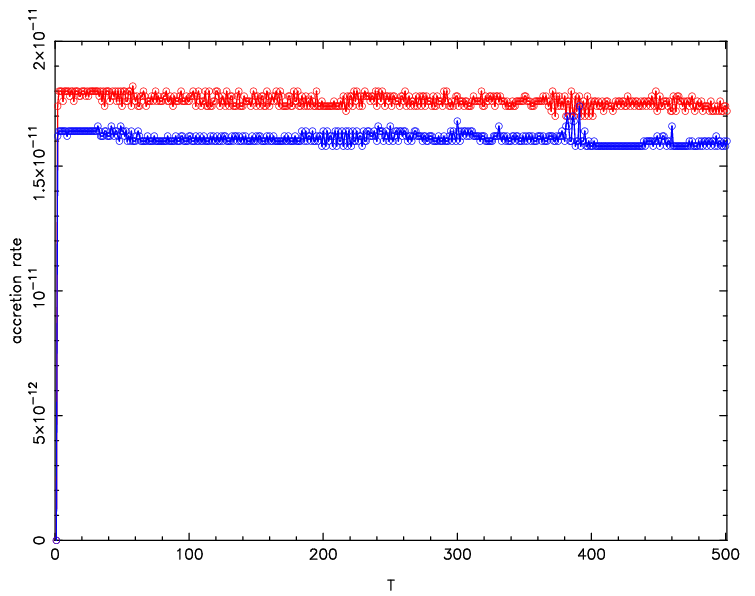}
   \end{tabular}
 \end{center}
    \caption{Accretion rate onto the inner boundary along with time (in unit of binary orbit) with the Newtonian (red line) and post-Newtonian (blue line) hydrodynamics. Left and middle show the results when $\nu=10^{-5}$ and $\nu=2\times10^{-4}$ with $a_0 =40 r_{\rm s}$, respectively. Right shows the results when $\nu=2\times10^{-4}$ with $a_0 =80 r_{\rm s}$.}
    \label{fig:figure4}
\end{figure*}

\begin{figure*}
 \begin{center}
   \begin{tabular}{cc}
      \includegraphics[width=0.25\textwidth]{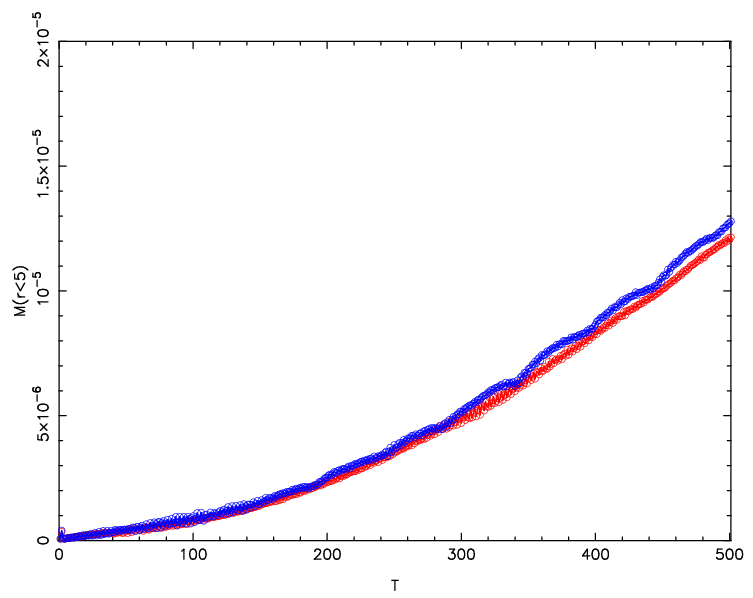}
      \includegraphics[width=0.25\textwidth]{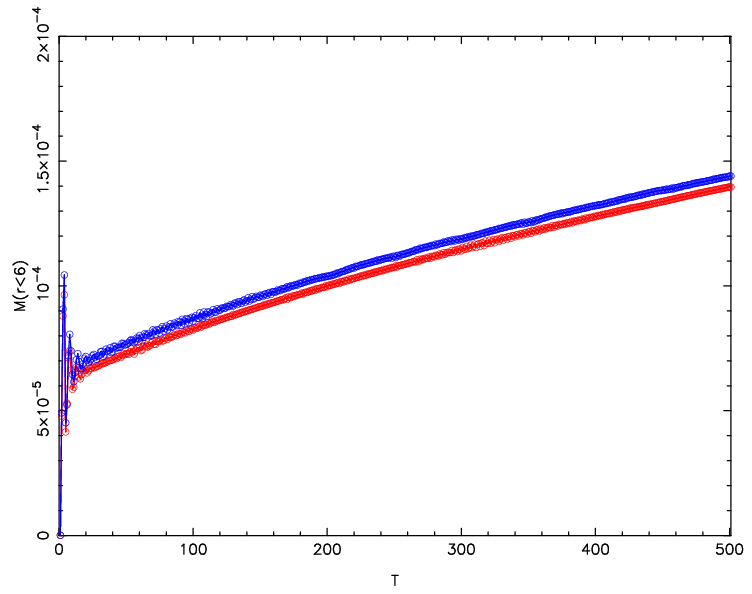}
      \includegraphics[width=0.25\textwidth]{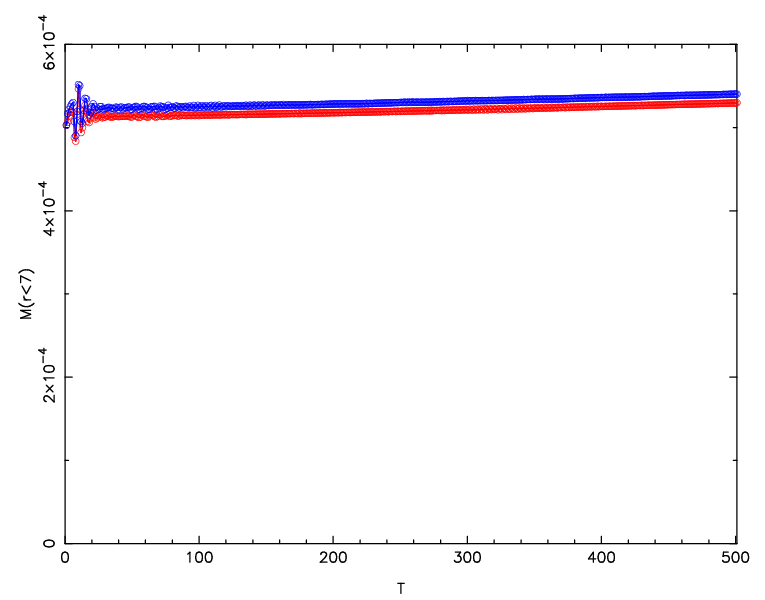}
      \includegraphics[width=0.25\textwidth]{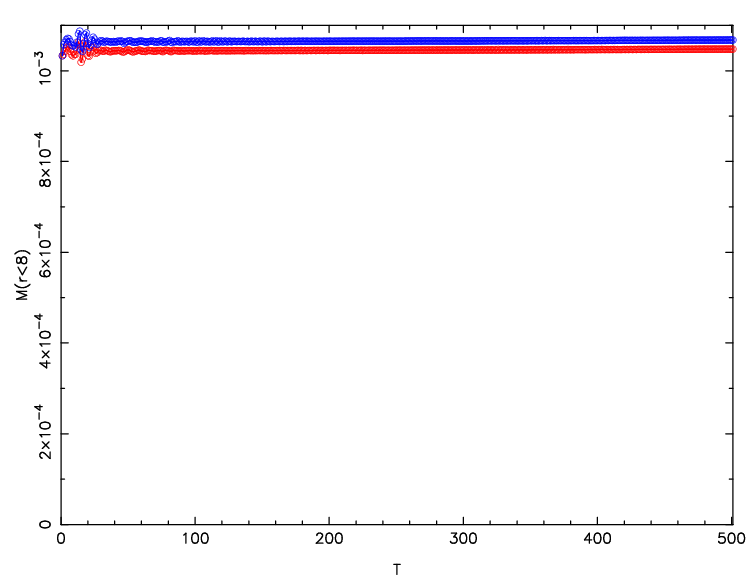}\\
      \includegraphics[width=0.25\textwidth]{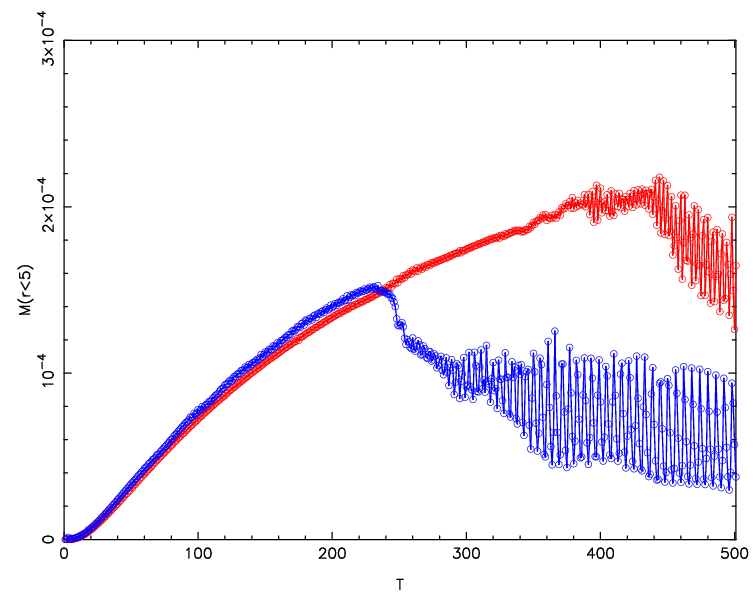}
      \includegraphics[width=0.25\textwidth]{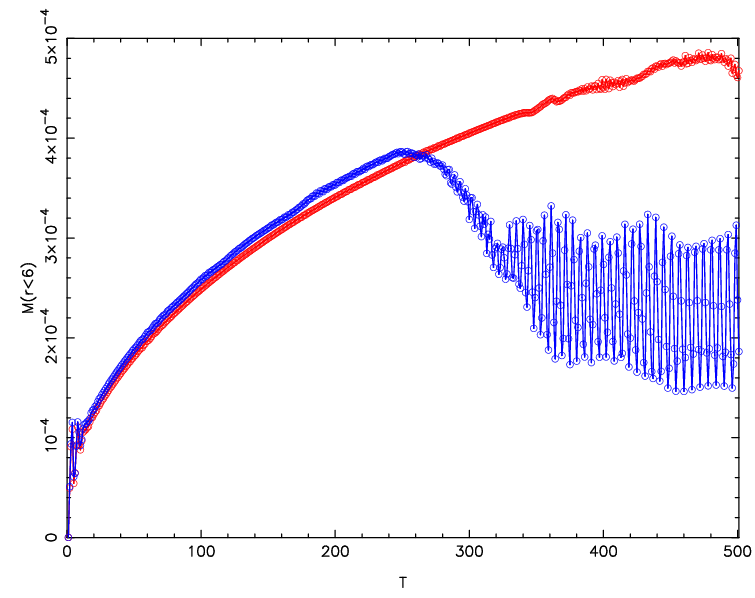}
      \includegraphics[width=0.25\textwidth]{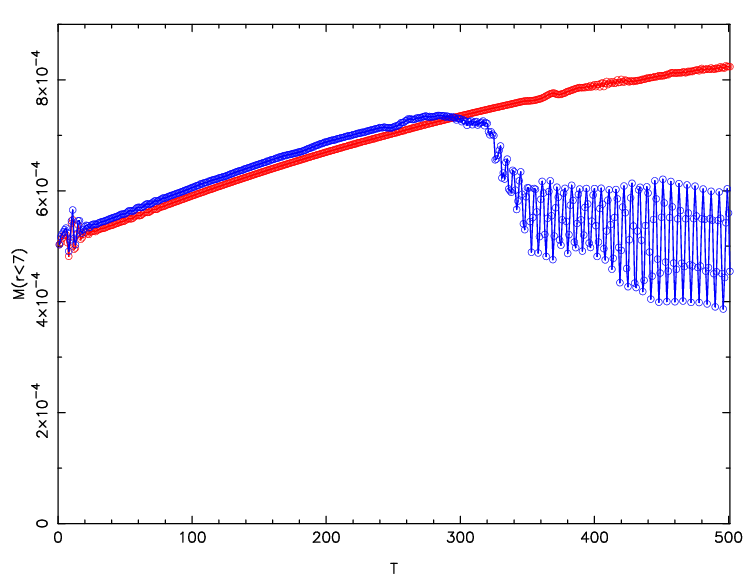}
      \includegraphics[width=0.25\textwidth]{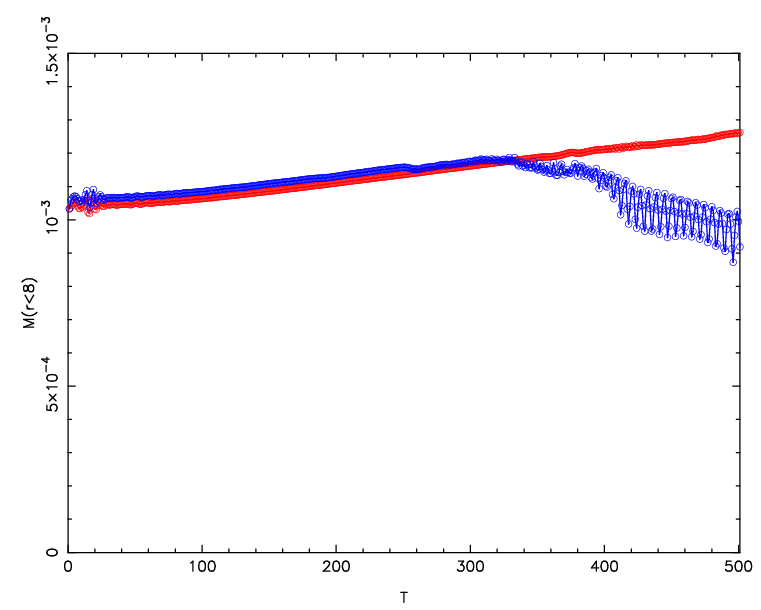}\\
      \includegraphics[width=0.25\textwidth]{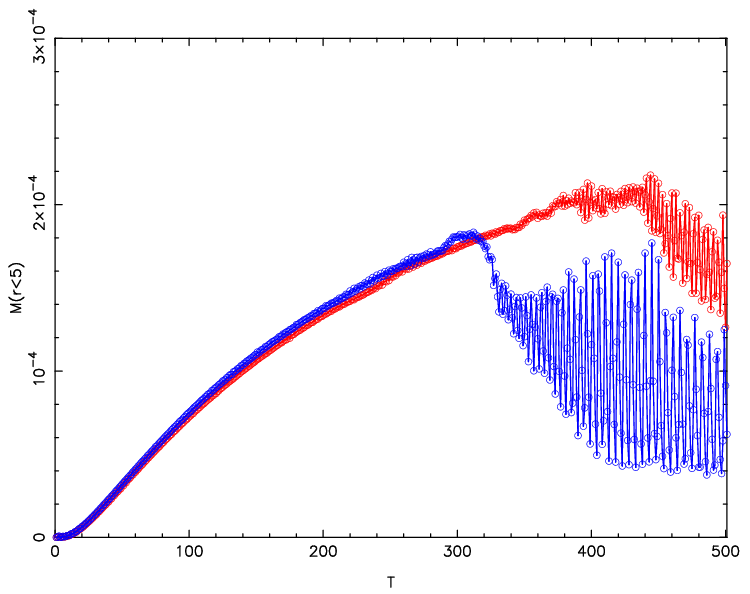}
      \includegraphics[width=0.25\textwidth]{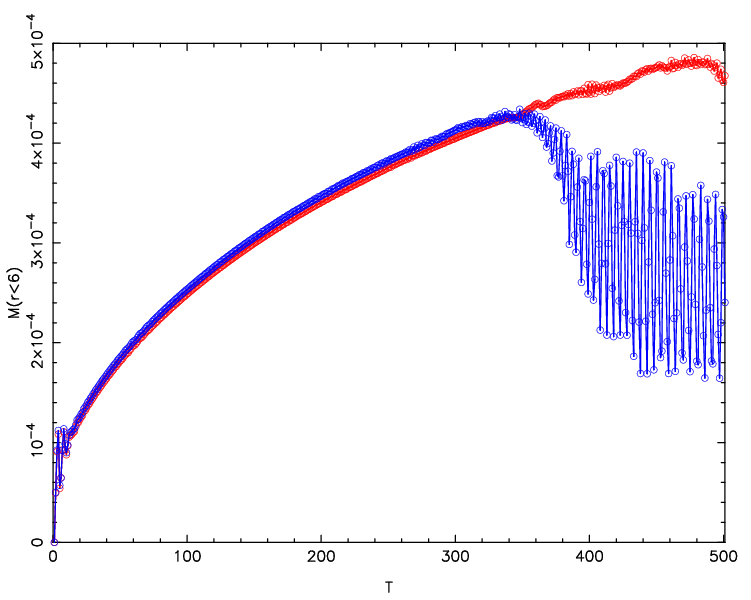}
      \includegraphics[width=0.25\textwidth]{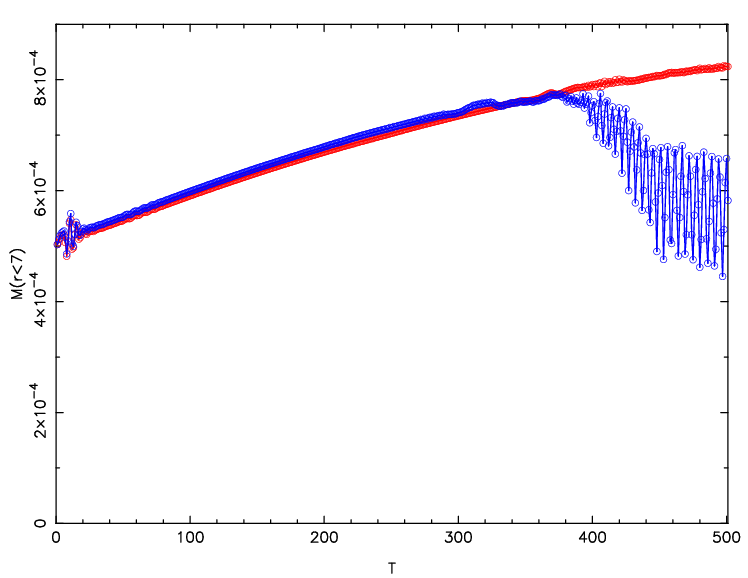}
      \includegraphics[width=0.25\textwidth]{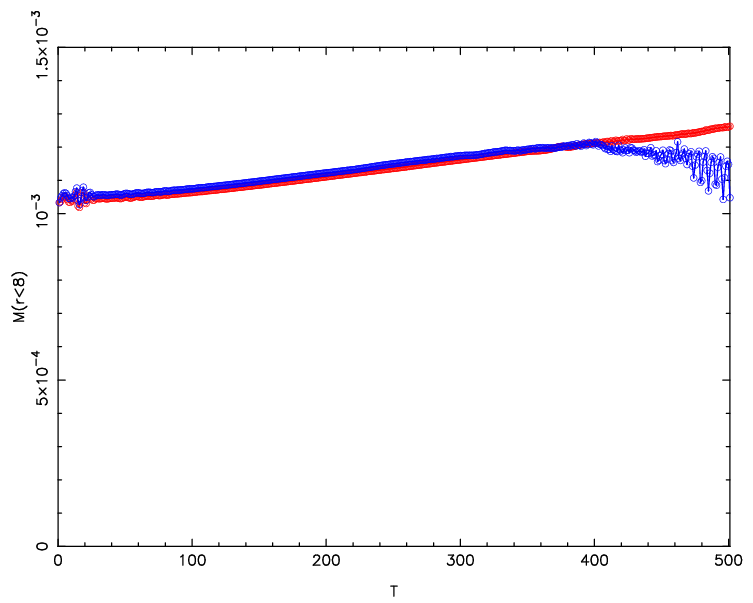}
   \end{tabular}
 \end{center}
    \caption{The enclosed mass within volumes of different radius with the Newtonian (red line) and post-Newtonian (blue line) hydrodynamics. Up and middle show the results when $\nu=10^{-5}$ and $\nu=2\times10^{-4}$ with $a_0 =40 r_{\rm s}$, respectively. Bottom shows the results when $\nu=2\times10^{-4}$ with $a_0 =80 r_{\rm s}$.}
    \label{fig:figure5}
\end{figure*}

\begin{figure*}
 \begin{center}
   \begin{tabular}{cc}
      \includegraphics[width=0.25\textwidth]{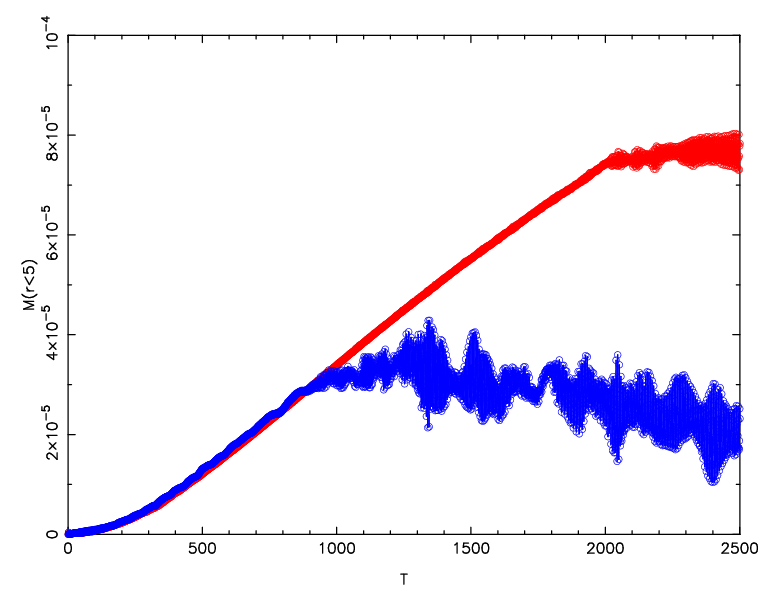}
      \includegraphics[width=0.25\textwidth]{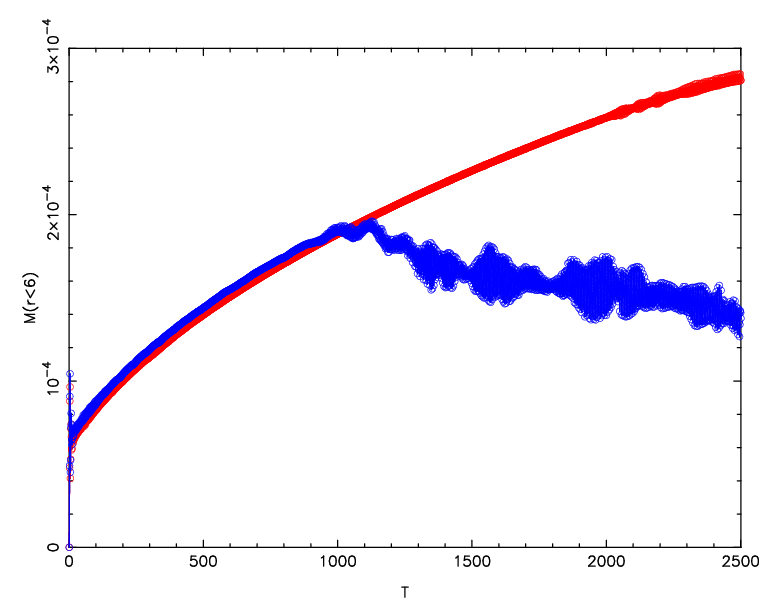}
      \includegraphics[width=0.25\textwidth]{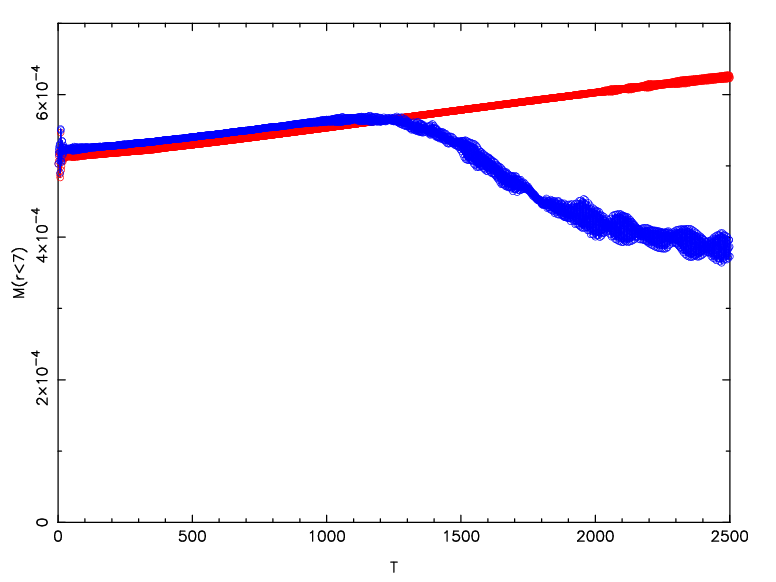}
      \includegraphics[width=0.25\textwidth]{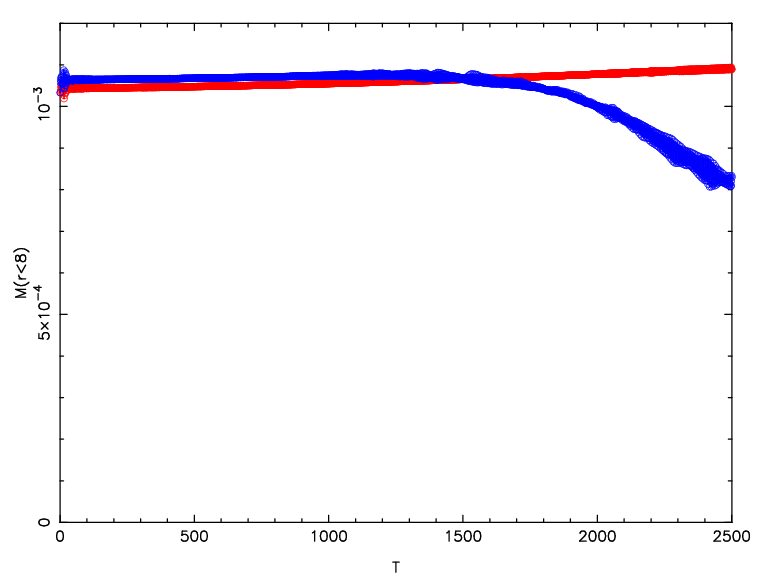}\\
      \includegraphics[width=0.25\textwidth]{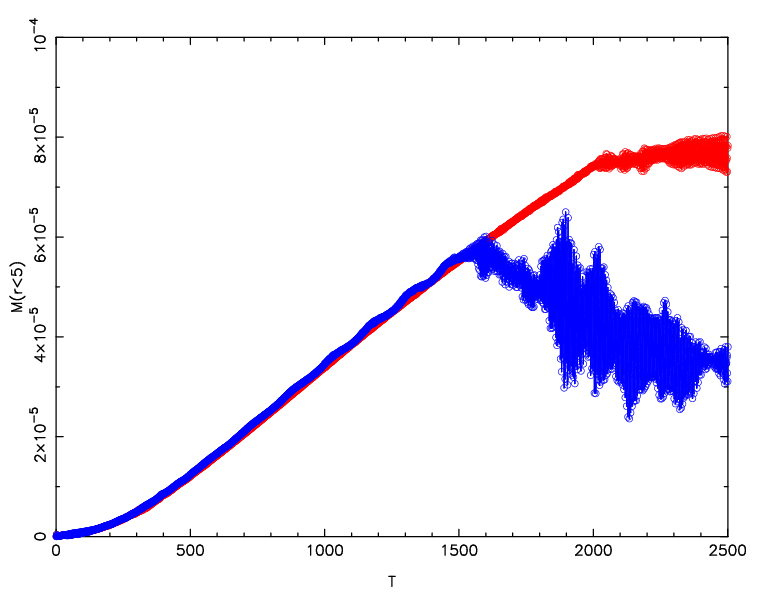}
      \includegraphics[width=0.25\textwidth]{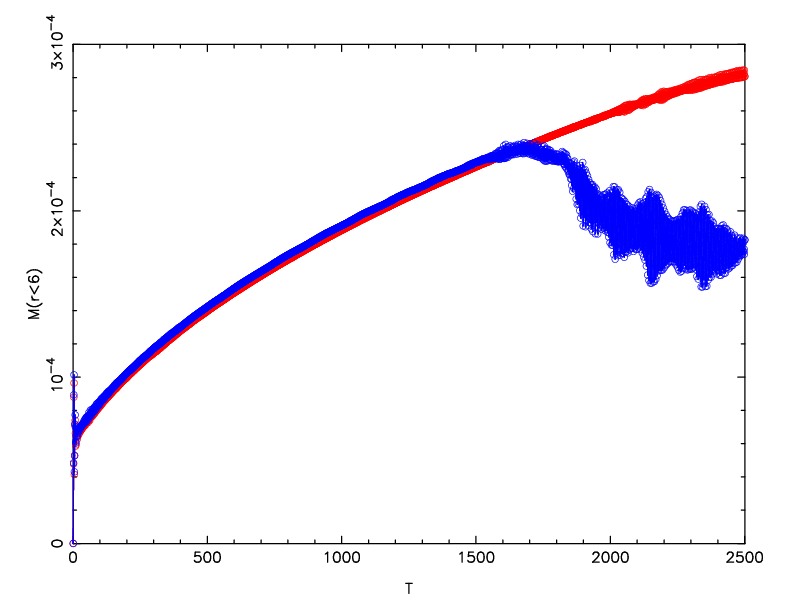}
      \includegraphics[width=0.25\textwidth]{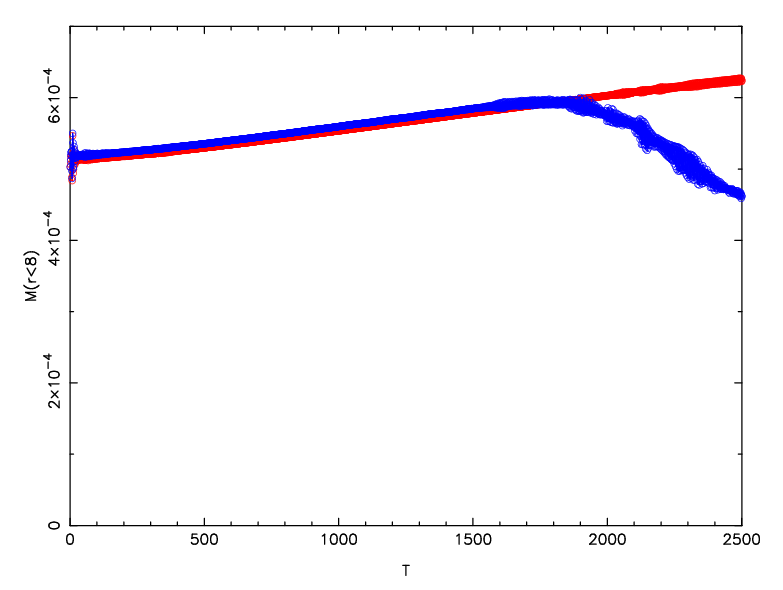}
      \includegraphics[width=0.25\textwidth]{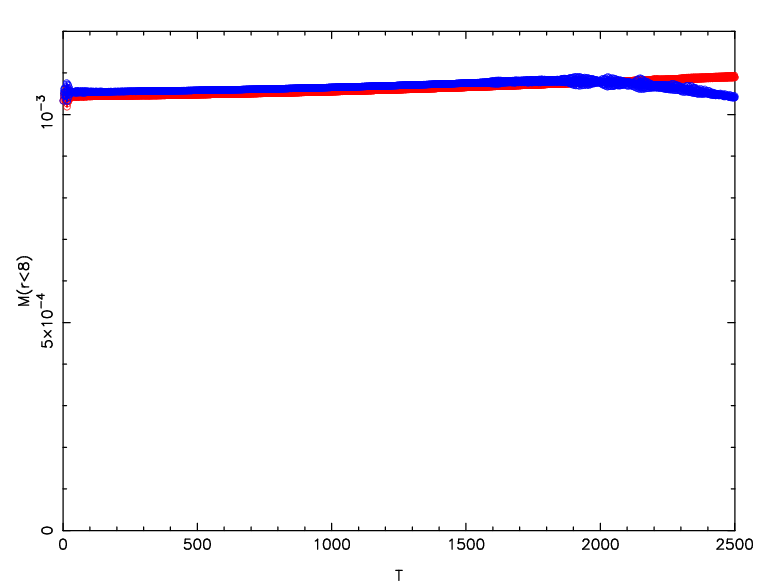}
   \end{tabular}
 \end{center}
    \caption{The enclosed mass within volumes of different radius with the Newtonian (red line) and post-Newtonian (blue line) hydrodynamics.Up and bottom represent the results from simulations when $\nu=10^{-5}$ with $a_0 =40 r_{\rm s}$ and $a_0 =80 r_{\rm s}$, respectively.}
    \label{fig:figure6}
\end{figure*}

\begin{figure*}
 \begin{center}
   \begin{tabular}{cc}
      \includegraphics[width=0.33\textwidth]{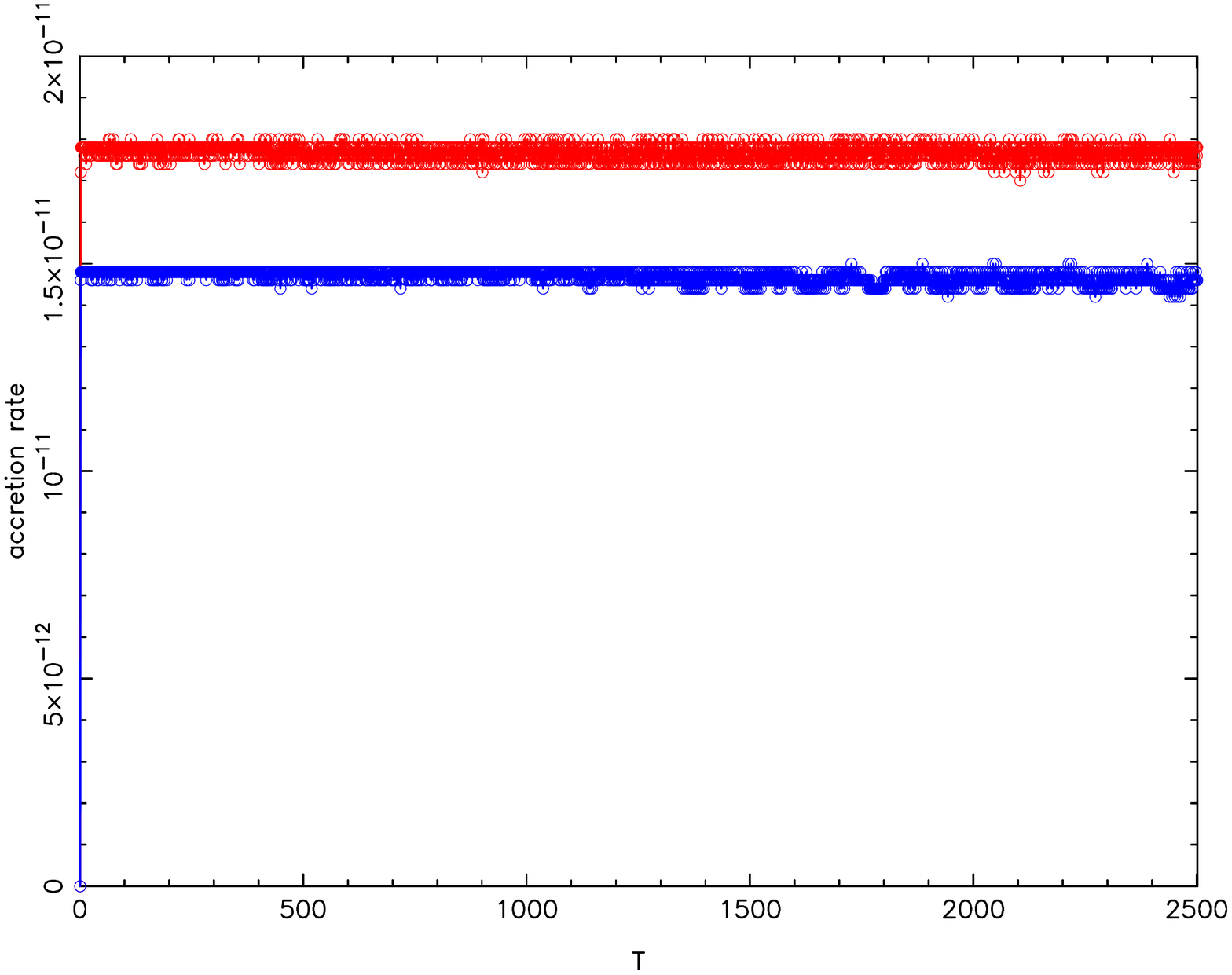}
      \includegraphics[width=0.33\textwidth]{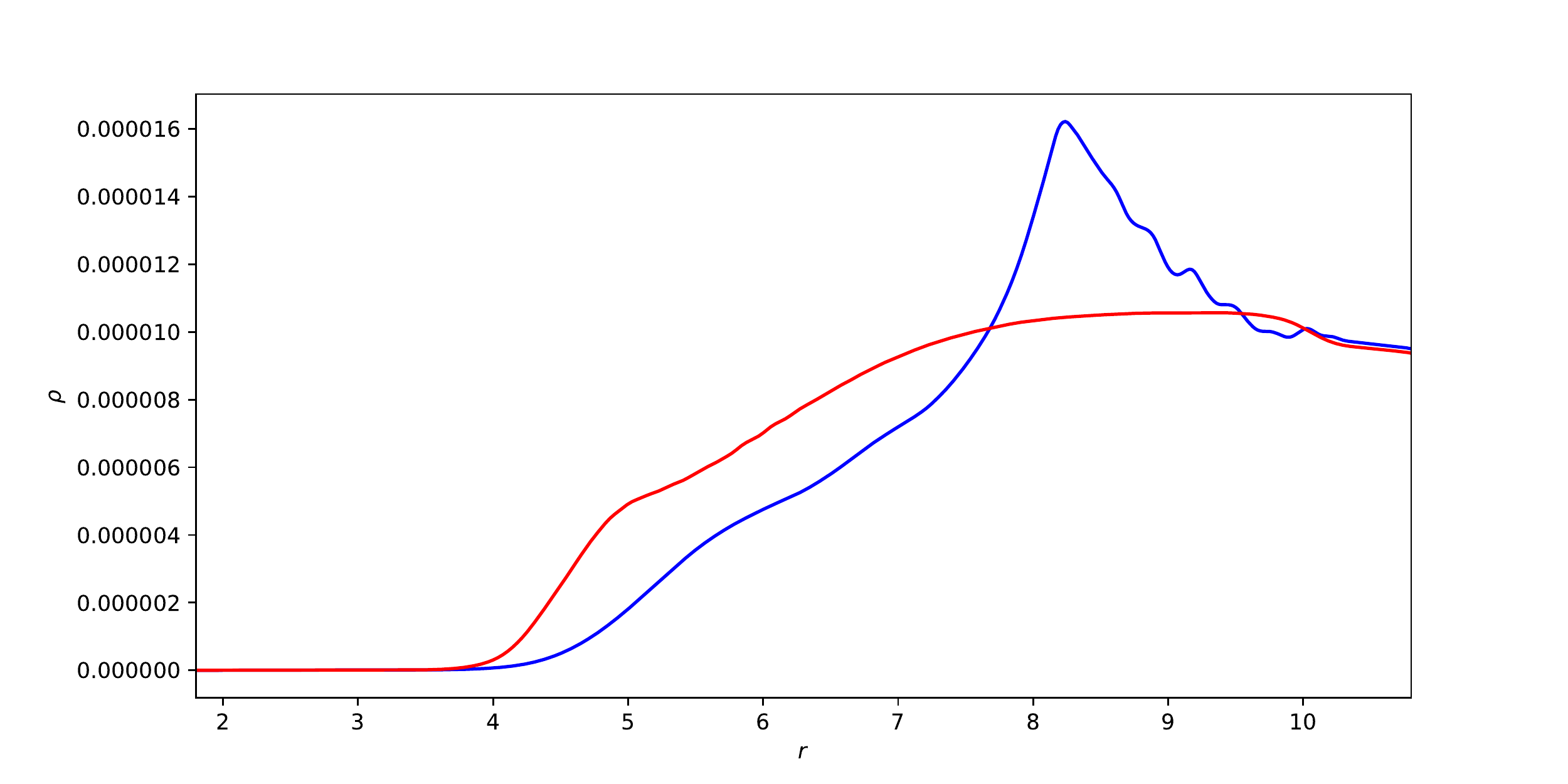}
      \includegraphics[width=0.33\textwidth]{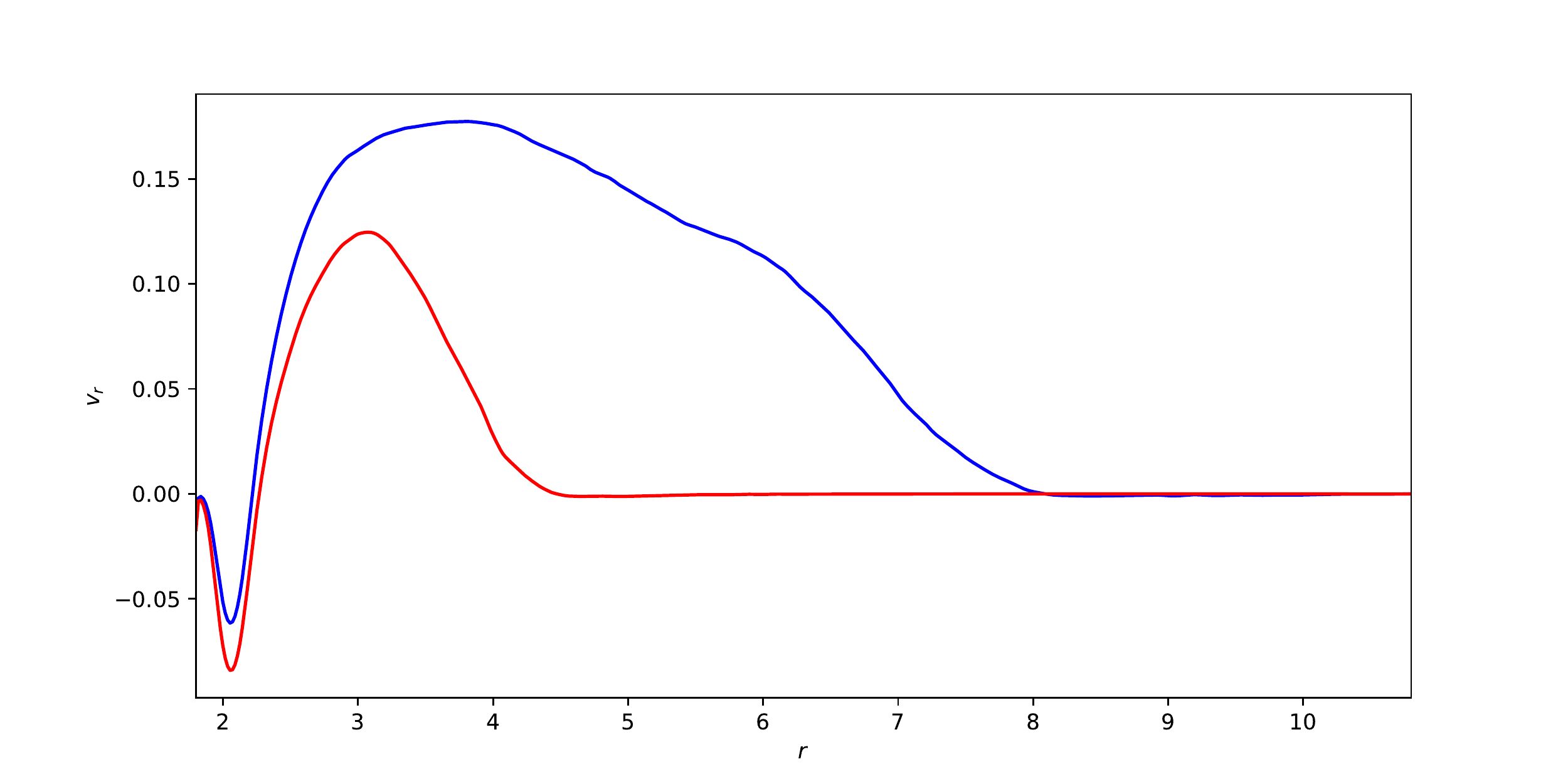}\\
      \includegraphics[width=0.33\textwidth]{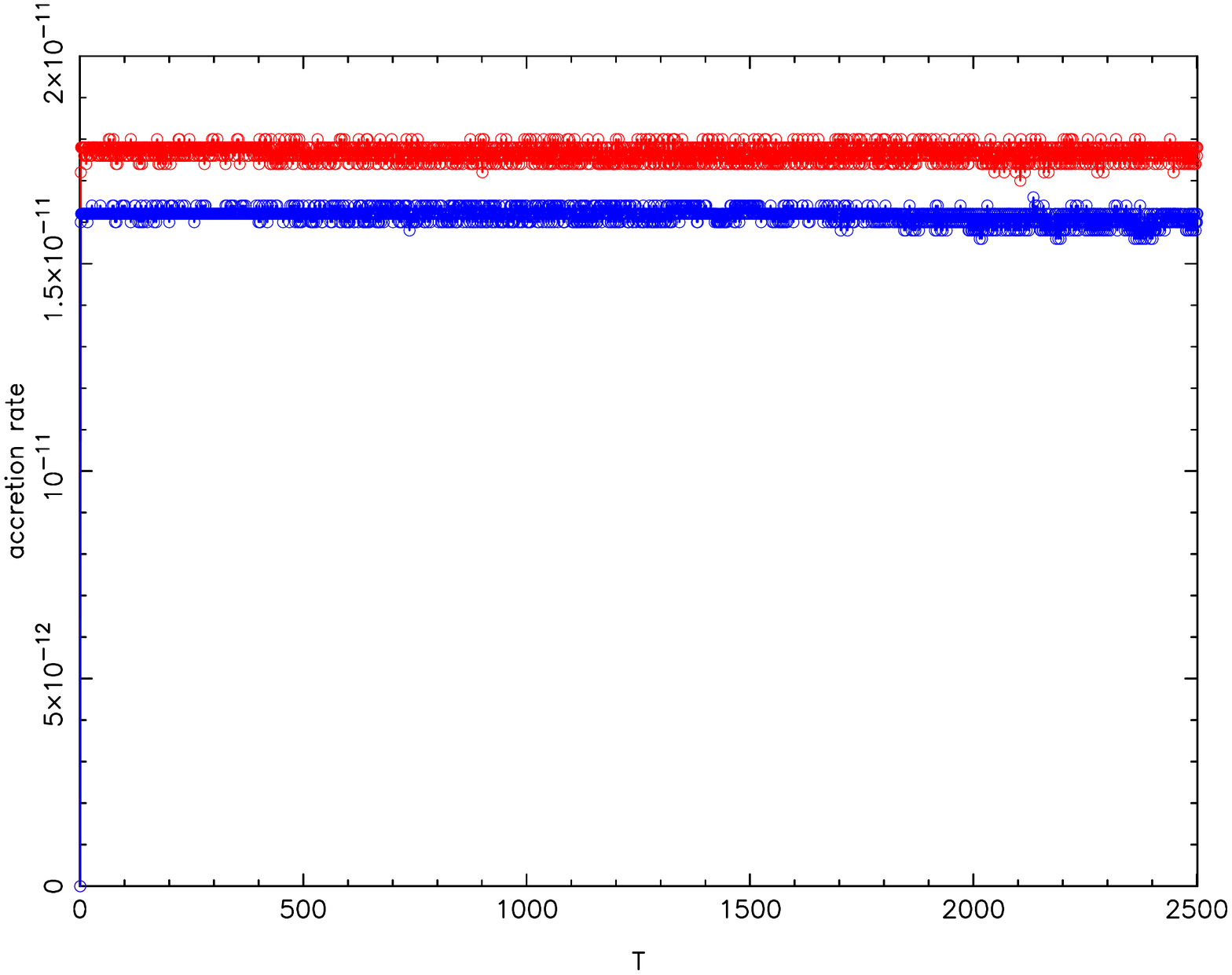}
      \includegraphics[width=0.33\textwidth]{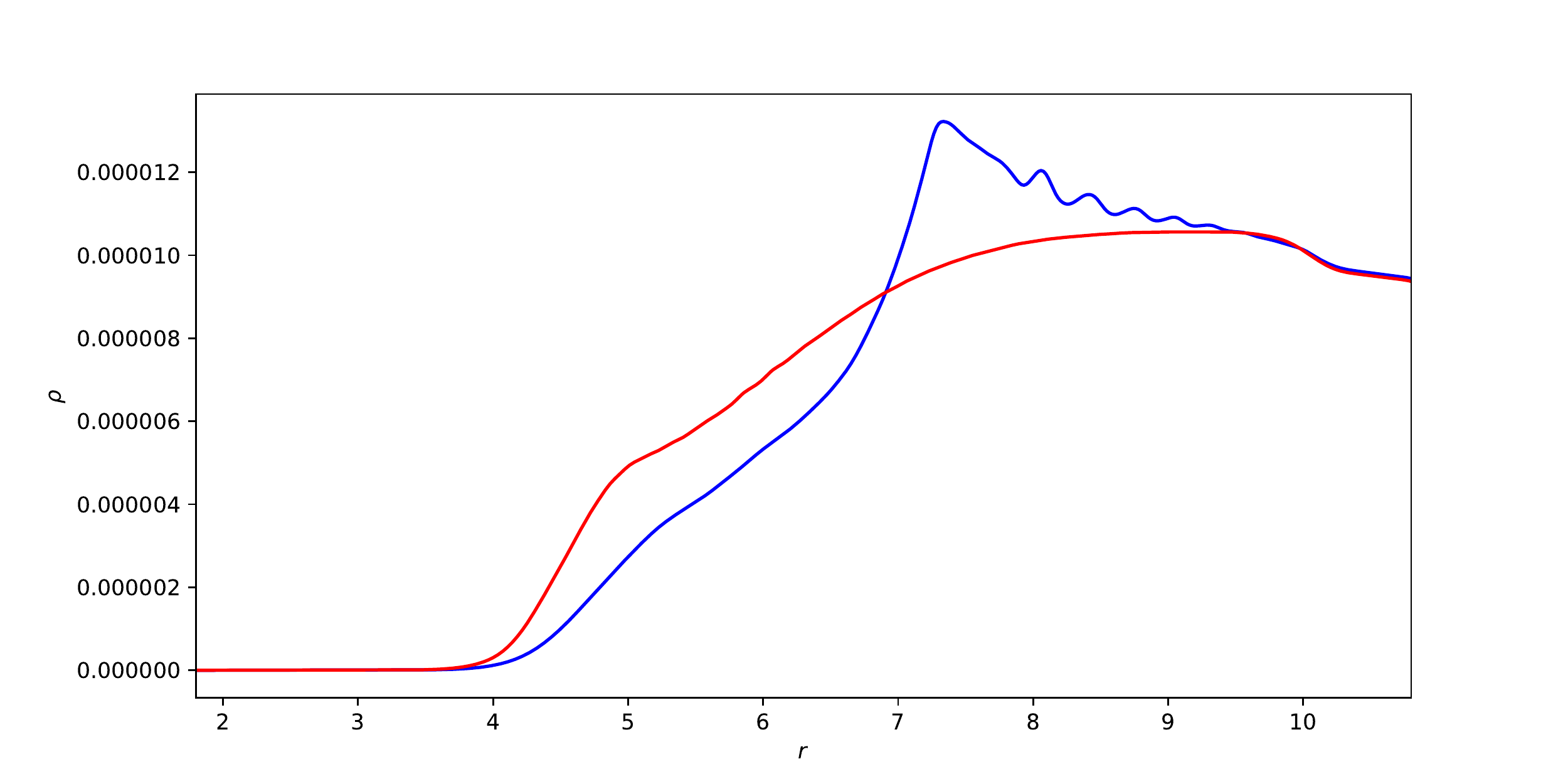}
      \includegraphics[width=0.33\textwidth]{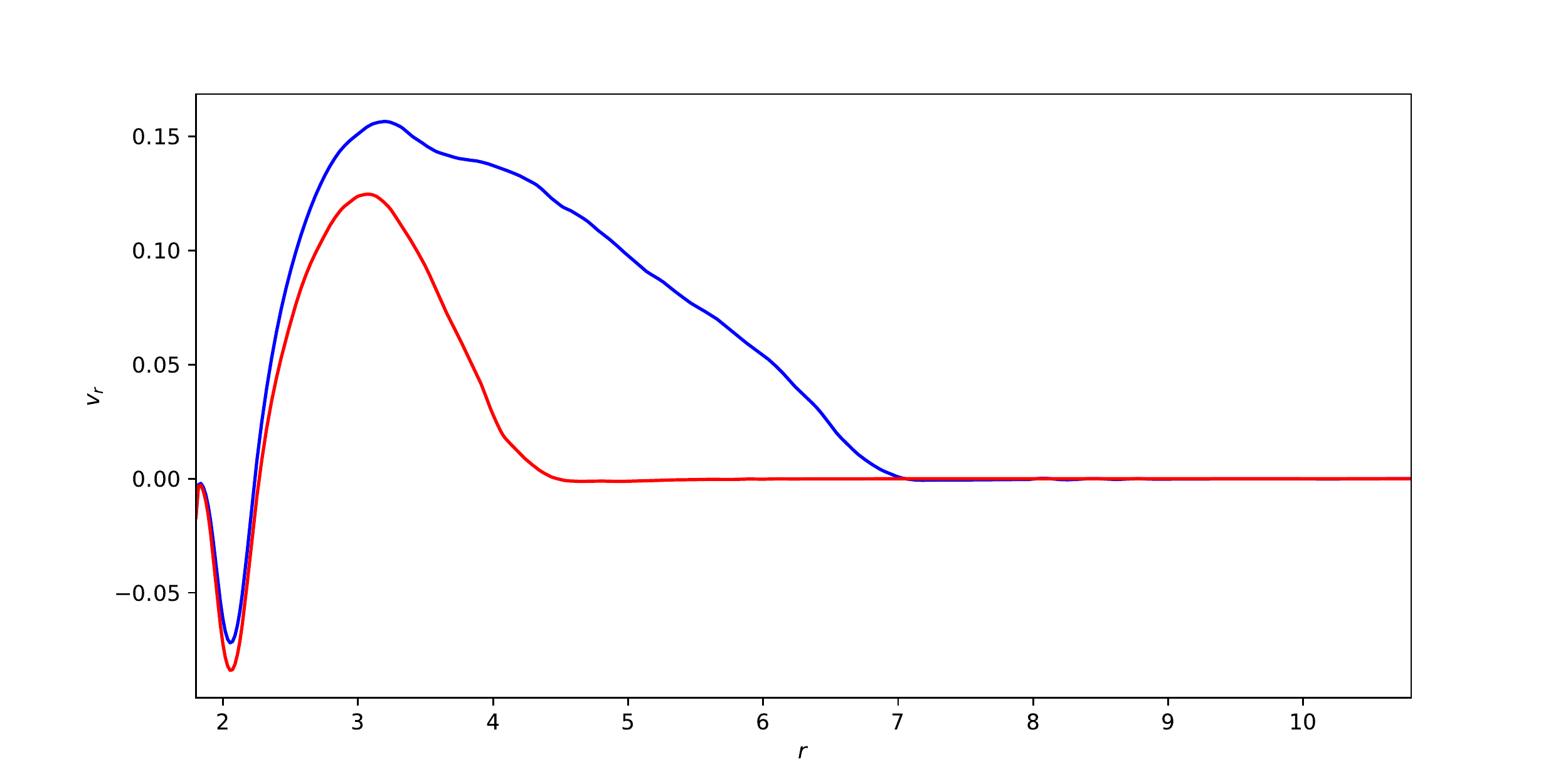}
   \end{tabular}
 \end{center}
    \caption{Left shows the accretion rate onto the inner boundary along with time in the Newtonian (red line) and post-Newtonian (blue line) regime. Middle shows the azimuthally averaged surface density along with radius which is averaged over 100 orbits from $2400$ to $2500$ orbits of the binary with the Newtonian (red line) and post-Newtonian (blue line) hydrodynamics. The azimuthally averaged radial velocity along with radius which is averaged over 100 orbits from $2400$ to $2500$ orbits of the binary with the Newtonian (red line) and post-Newtonian (blue line) hydrodynamics is shown in the right panel. Up and bottom represent the results from simulations when $\nu=10^{-5}$ with $a_0 =40 r_{\rm s}$ and $a_0 =80 r_{\rm s}$, respectively.}
    \label{fig:figure7}
\end{figure*}

\begin{figure*}
 \begin{center}
   \begin{tabular}{cc}
      \includegraphics[width=0.5\textwidth]{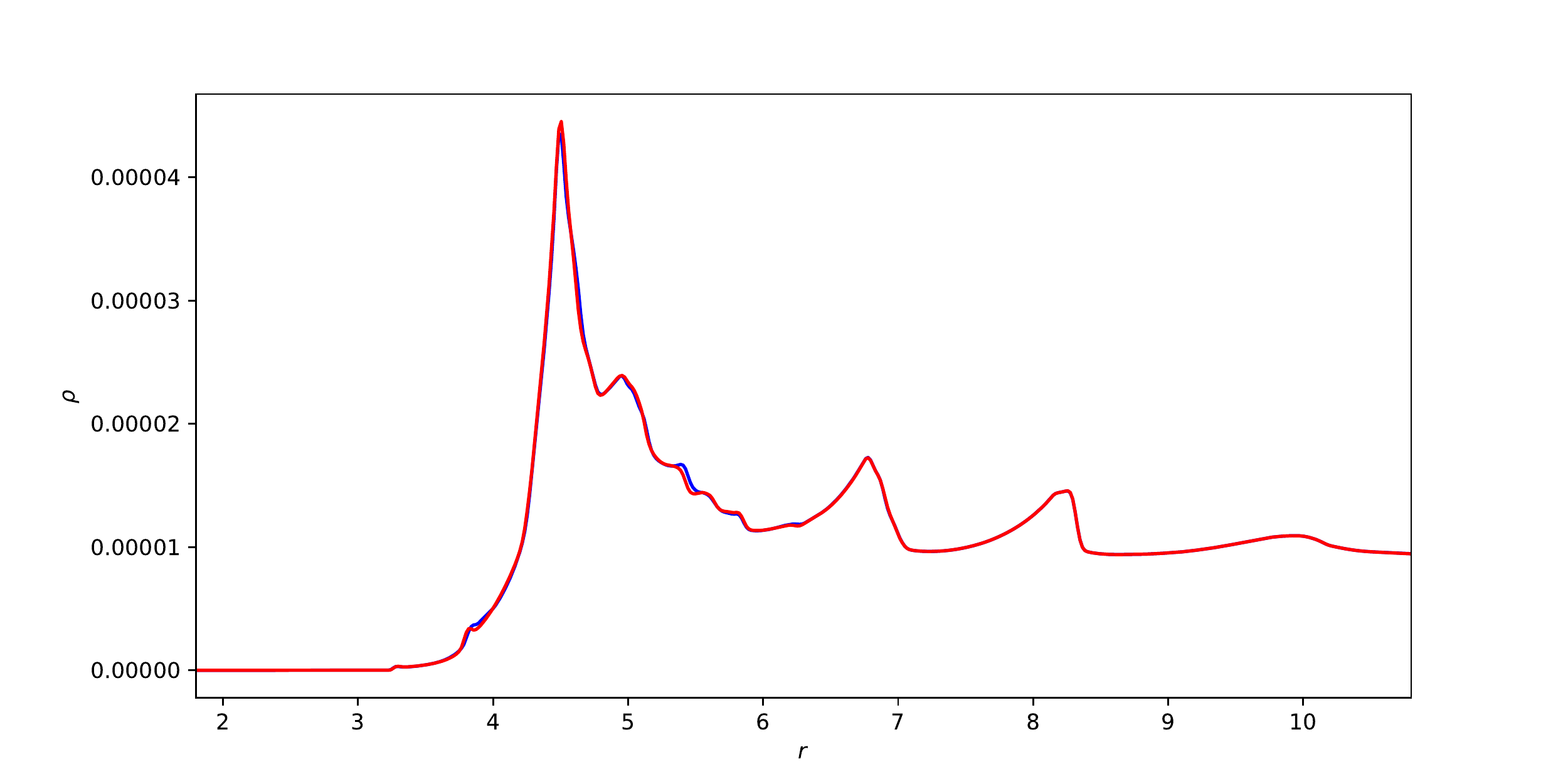}
      \includegraphics[width=0.5\textwidth]{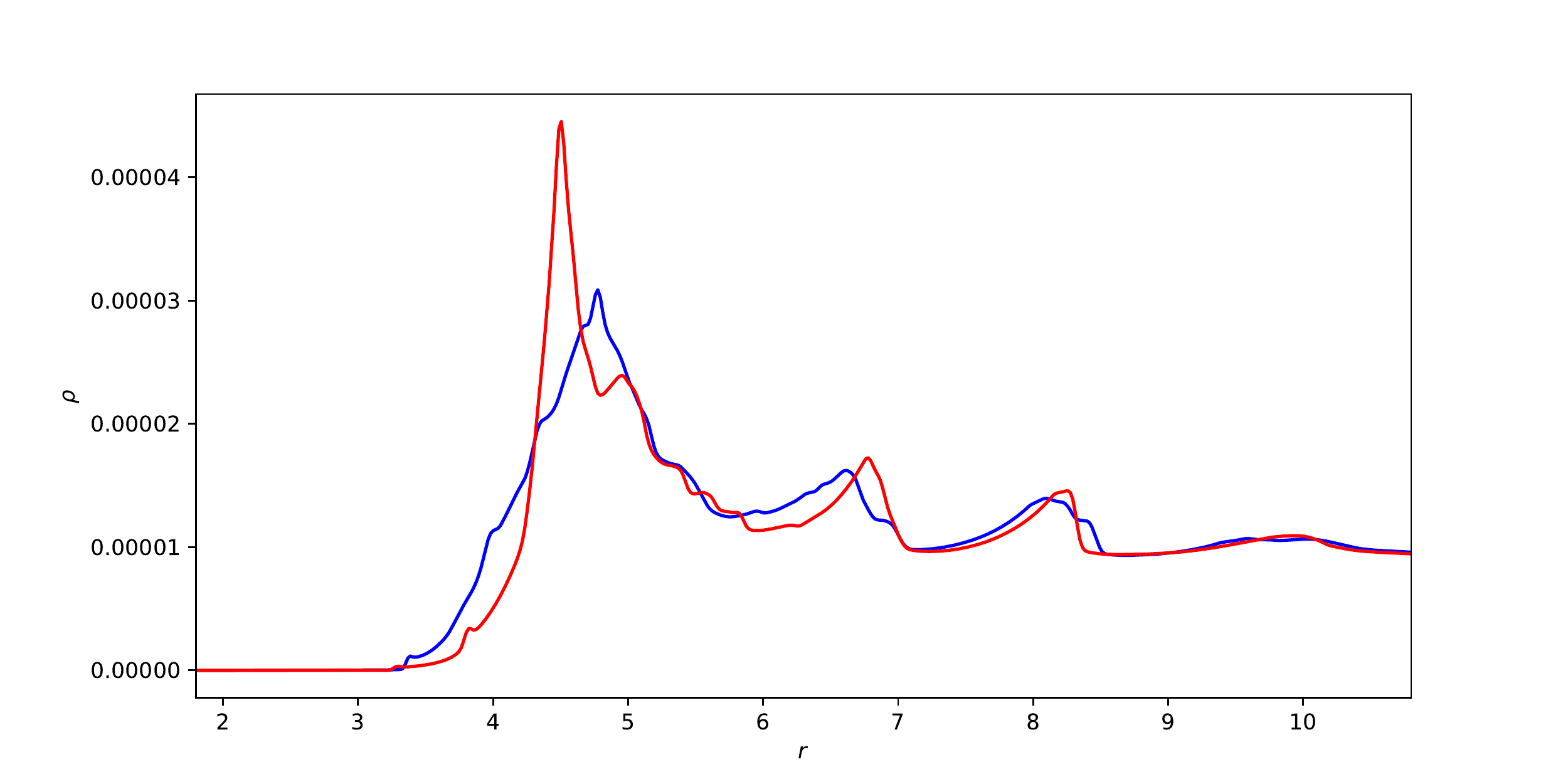}
   \end{tabular}
 \end{center}
    \caption{Red and blue lines show the azimuthally averaged surface density along with radius at 10 orbit of the binary with the Newtonian (red line) and post-Newtonian (blue line) hydrodynamics when $\nu=10^{-5}$, respectively. Left and right represent the simulations with the binary separation of $a=20000r_{\rm s}$ and $a=40r_{\rm s}$, respectively.}
    \label{fig:figure8}
\end{figure*}

\section{Results and Analysis}
Since the disk has a finite mass and radius, the disk should evolve for enough time to overcome the initial transient \citep{Ragusa16,Terquem17}. We evolve the disk with 500 orbits of the binary which is about two times (half) of the timescale needed by the initial inner edge of the disk to reach about $2.5a_0$ ($2a_0$) when $\nu=2\times10^{-4}$ according to the viscous timescale $\tau_{\rm visc} \sim \frac{\Delta r^2}{\nu}$ \citep{Durmann15}, and the simulated time is less than the merger timescale for $a_0 =80 r_{\rm s}$ (about 730 orbits). The simulations with $\nu_1=10^{-5}$ are taken as comparison with that of $\nu=2\times10^{-4}$ in order to demonstrate the importance of viscosity on the evolution of the disk in the PN hydrodynamics, and the results from the simulations with $a_0 =40 r_{\rm s}$ compared with that with $a_0 =80 r_{\rm s}$ could illustrate the effect of the binary separation on the PN hydrodynamics. Figure.~\ref{fig:figure1} shows the density distribution of the disk after about 500 orbits of the binary when $\nu=2\times10^{-4}$ with $a_0 =40 r_{\rm s}$ and $a_0 =80 r_{\rm s}$ while Figure.~\ref{fig:figure2} presents the azimuthally averaged surface density which is averaged over 100 orbits from $400$ to $500$ orbits of the binary. From Figure.~\ref{fig:figure2}, we observe that the gap cleared by the binary of equal mass in the PN regime is wider than that in the Newtonian regime when $\nu=2\times10^{-4}$ and the disk around the binary tends to have a larger region with positive radial velocity in the PN regime than that in the Newtonian regime from Figure.~\ref{fig:figure3} when $\nu=2\times10^{-4}$. Figure.~\ref{fig:figure1} also shows that an eccentric cavity is developed in each simulation, which has been widely observed in other works \citep{Papaloizou01,D'Angelo06,Kley06,MacFadyen08,D'Orazio13,Farris14,Dunhill15,Ragusa16} and could be interpreted as being caused by the unstable growth of spontaneous deviations of the gaseous element from circular motion \citep{Lubow91}.

Figure.~\ref{fig:figure4} shows the accretion rate onto the inner boundary along with time with different uniform kinematic viscosity. The accretion rate tends to be lower in the PN hydrodynamics than that in the Newtonian hydrodynamics. The enclosed mass within volumes of different radius is presented in Figure.~\ref{fig:figure5}, showing that the amount of mass within different radius in the PN hydrodynamics is larger along with time than that in the Newtonian hydrodynamics when $\nu=10^{-5}$ from the up panel of Figure.~\ref{fig:figure5} while that it's reverse in the middle and bottom panel of Figure.~\ref{fig:figure5} when $\nu=2\times10^{-4}$ after a specific time. With the initial disk in quasiequilibrium and after about 500 binary orbits, the difference of the distribution of the azimuthally averaged surface density along with radius between the PN hydrodynamics and Newtonian hydrodynamics is minor in the case of a relatively small uniform kinematic viscosity of $\nu=10^{-5}$. But with a relatively large uniform kinematic viscosity of $\nu=2\times10^{-4}$, the difference is obvious in the middle and bottom panel of Figure.~\ref{fig:figure5}. The difference between the two cases may be due to the simulational time span. For this reason, we run the low-viscosity simulation with $\nu=10^{-5}$ five times longer so that it is run for about the same factor times $\tau_{visc}$ as currently the high-viscosity run with $\nu=2\times10^{-4}$, the results are presented in
Figure.~\ref{fig:figure6} and Figure.~\ref{fig:figure7}. The enclosed mass within volumes of different radius shown in Figure.~\ref{fig:figure6} appears to have the same properties as that in the middle and bottom panel of Figure.~\ref{fig:figure5}. Results in Figure.~\ref{fig:figure7} with the accretion rate, the azimuthally averaged surface density and radial velocity averaged over 100 orbits from $2400$ to $2500$ orbits of the binary both in Newtonian and PN regime, present features similar to that in Figure.~\ref{fig:figure2},
Figure.~\ref{fig:figure3} and Figure.~\ref{fig:figure4}. As a consequence, it means that a wider gap resulting from PN effect emerges at all viscosities and isn't restricted to the high-viscosity case.

The gap shown in Figure.~\ref{fig:figure1} and the middle and right panel of Figure.~\ref{fig:figure2} seems to be wider in the PN regime and might present much more unique observable signatures of the continuum emission in such binary-disk system \citep{Gultekin12,D'Orazio13,Yan15,Ryan17,Tang18}. The effect of gap forming around SMBBH on the thermal spectrum ranges from a sharp exponential cut-off in the IR-optical-UV spectral energy distribution (SED) due to the emission which could be truncated blueward of the wavelength corresponding to the temperature of the innermost circumbinary disk edge \citep{Gultekin12,D'Orazio13} (only the circumbinary disk is considered), to a notch arising from the emergence of the gap between the tidal radii of the mini-disk around each SMBH and the truncation radius of the circumbinary disk \citep{Yan15,Ryan17,Tang18} (both the mini-disks and the circumbinary disk are taken into consideration). Since the gap produced by the binary is wider in the PN regime than that in the Newtonian regime, a much more obvious notch in the SED may be emerge in the PN regime and we will study the feature of the SED resulting from binary-disk interaction in the PN hydrodynamics with the inner zone metric and the near zone metric in detail in future.

In this work with a fixed binary separation, we simulate the binary-disk interaction with 500 orbits in order to investigate the PN effect on the dynamics of the circumbinary disk. The time needed by the binary black hole to merge due to gravitational radiation with initial separation of 40$R_s$ is about 128 orbits and the binary with initial separation of 80$R_s$ will merge after about 730 orbits. These two sets of simulations with different fixed separation show similar properties in the PN hydrodynamics with the minor difference in the enclosed mass of a given radius, meaning that the effect of PN dynamics on the disk depends on the binary separation and the PN effect will increase as the separation of the binary decreases. Although we fixed the SMBBH separation for numerical convenience, in reality, the binary would inspiral during the simulated time. During inspiral of the SMBBH, the PN effect on the dynamics of the circumbinary disk will become more and more important as the separation of the binary decreases and may have a major impact on the size and shape of the gap. In future, we will simulate the SMBBH-disk interaction using PN hydrodynamics with both the inner zone metric and the near zone metric in order to study the dynamics of the mini-disk around each SMBH and the circumbinary disk during inspiral.

When the separation of the binary is extremely large, we conduct two additional simulations using both Newtonian and PN hydrodynamics with the same parameters in unit in Section 2 except that the separation is changed to $a_0 \approx 20000 r_{\rm s}$. The azimuthally averaged surface density after ten orbits is shown in the left panel of Figure.~\ref{fig:figure8} when $a_0 \approx 20000 r_{\rm s}$ with the initial disk whose inner radius is $r=3$ (gravitational torque is strong with such small radius) and $\nu=10^{-5}$, and the right panel presents the azimuthally averaged surface density when $a_0 = 40 r_{\rm s}$ after ten orbits. It shows that the effect of the PN hydrodynamics becomes weak with such large binary separation $a_0 \approx 20000 r_{\rm s}$. When the binary separation becomes extremely large, the PN hydrodynamics returns to the Newtonian hydrodynamics.

As the separation of the SMBBH decreases by interacting with the disk around them, the relativistic effects of the spacetime of the SMBBH will become significant. Hydrodynamical simulations about the binary-disk interaction with full numerical relativity are much time expensive, following the stage near merger in recent works \citep{Farris12,Giacomazzo12}. In the case of a relatively large separation, Newtonian hydrodynamics could account for the evolution. At a small binary separation and with the condition that the binary will take a long time to merge and the PN hydrodynamics could describe the evolution of the system precisely, method in this work, with the advantages that the PN hydrodynamics is more precise than the Newtonian hydrodynamics and is faster in computation than the full numerical relativity, could be used as the method to investigate the binary-disk interaction in the transitional stage between the stage where Newtonian regime dominates the binary-disk evolution given that the separation of the binary is sufficiently large and the stage where the full numerical relativity should be adopted in order to simulate the binary-disk interaction precisely provided that the separation of the binary is sufficiently small.


\section{CONCLUSIONS} \label{discuss}
The formation of SMBBH could result from the merger of two galaxies. The gravitational interaction between the SMBBH and the background stars and gaseous disk shrinks the binary separation until gravitational radiation becomes the dominant source of extraction of angular momentum and energy of the binary.

In this work, we investigate the effect of 1PN on the accretion disk surrounding SMBBH of equal mass whose separation is small enough. By adopting the PN hydrodynamics with the near zone of 1PN expansion of the binary black hole metric, we simulate the evolution of accretion disk around the binary. The simulational results show similar properties and seem to be independent of disk viscosity provided that the time span in simulations has about the same factor times the viscous timescale associated with different viscosities. The behavior of the evolution of the circumbinary disk with PN hydrodynamics could have qualitative difference in contrast to that with Newtonian hydrodynamics. The accretion rate onto the inner boundary is lower in the PN hydrodynamics than that in the Newtonian hydrodynamics and the amount of mass within different radius in the PN hydrodynamics tends to be lower along with time, resulting that the circumbinary disk around SMBBH of equal mass tends to have a wider gap when adopting PN hydrodynamics compared with Newtonian hydrodynamics and meaning that the disk may show much more unique observable signatures of the continuum emission in the PN regime. This work investigates the SMBBH-disk interaction with each black hole mass of $M_1 =5\times10^8 M_{\odot}$ and binary separation of $a_0 =40 r_{\rm s}$ and $a_0 =80 r_{\rm s}$ using Newtonian hydrodynamics and PN hydrodynamics. PN hydrodynamical simulations with parameters of other separation and mass of the SMBBH could be investigated with the same method in this work. To account for the evolution of the disk near each black hole of the binary with 1PN accurately, the inner zone metric should be considered and we will conduct such work in future.

\textbf{Acknowledgements}

I am very grateful to the anonymous referee for insightful comments that improved this work. This work is supported by the National Science Foundations of China (U1931203).

\textbf{Data Availability}

The data underlying this article will be shared on reasonable request to the corresponding author.



\bsp	
\label{lastpage}
\end{document}